\documentclass[review,onefignum,onetabnum]{article}
\usepackage{latexsym}
\usepackage{amssymb,amsbsy,amsmath,amsfonts,amssymb,amscd,amsthm}
\usepackage{graphicx}
\usepackage{hyperref}
\usepackage{mathtools}
\usepackage{pdfsync}
\usepackage{epsfig,epstopdf}
\usepackage{xcolor}
\newcommand{\rev}[1]{#1}
\usepackage{hyperref}
\usepackage[title]{appendix}
\usepackage{verbatim}
\usepackage{lipsum}
\usepackage{float}
\usepackage[font=small]{caption}
\usepackage{subcaption}
\usepackage[makeroom]{cancel}
\usepackage{booktabs}
\usepackage{braket}
\usepackage{tikz}

\newtheorem{rmrk}[]{Remark}

\DeclarePairedDelimiter {\abs}{\lvert}{\rvert}

\newcommand{\R}{\mathbb{R}}
\newcommand{\N}{\mathbb{N}}
\newcommand{\Z}{\mathbb{Z}}
\newcommand{\di}{\,\mathrm{d}}
\newcommand{\pt}{\partial_t}
\newcommand{\px}{\partial_x}

\newcommand{\pxx}{\partial_{xx}^2}

\newcommand{\beq}{\begin{equation}}
	\newcommand{\eeq}{\end{equation}}
\newcommand{\beqa}{\begin{eqnarray}}
	\newcommand{\eeqa}{\end{eqnarray}}

\newcommand{\btwo}{\tilde{\beta}}
\newcommand{\bthree}{{\beta}}

\providecommand{\keywords}[1]
{
	\small	
	\textbf{\textit{Keywords---}} #1
}

\renewcommand{\vec}{\boldsymbol}
\renewcommand{\epsilon}{\varepsilon}

\numberwithin{equation}{section}
\numberwithin{prpstn}{section}
\numberwithin{ass}{section}
\numberwithin{rmrk}{section}

\title{The role of viral dynamics and infectivity in models of oncolytic virotherapy for tumours with different motility\thanks{Corresponding author: David Morselli (d.morselli@ucl.ac.uk)\\
This research was partially supported by the Italian Ministry of Education, University and Research (MIUR) through the \rev{``Dipartimenti di Eccellenza'' Programme (2018-2022) -- Dipartimento di Scienze Matematiche ``G. L. Lagrange''}, Politecnico di Torino (CUP: E11G18000350001). MED and DM are members of GNFM (Gruppo Nazionale per la Fisica Matematica) of INdAM (Istituto Nazionale di Alta Matematica). FF acknowledges support from the Australian Research Council (ARC) via the Discovery Project (DP) DP230100485. We also acknowledge the support of the Australian National Health and Medical Research Council, through grant NHMRC IDEAS 2013058. Part of this work was performed on the OzSTAR national facility at Swinburne University of Technology. The OzSTAR program receives funding in part from the Astronomy National Collaborative Research Infrastructure Strategy (NCRIS) allocation provided by the Australian Government, and from the Victorian Higher Education State Investment Fund (VHESIF) provided by the Victorian Government.
}}
\author{David Morselli,\thanks{Department of Mathematics, University College London, 25 Gordon Street, London WC1H 0AY, United Kingdom}  \thanks{Department of Mathematics, School of Science, Computing and Emerging Technologies, Swinburne University of Technology, John St, 3122, Hawthorn, VIC, Australia} 
\and
Federico Frascoli,\footnotemark[2] 
\and
Marcello Edoardo Delitala\thanks{Department of Mathematical Sciences ``G. L. Lagrange'', Politecnico di Torino, Corso Duca degli Abruzzi 24, 10129 Torino, Italy}}

\begin{document}
	
\maketitle
	
\begin{abstract}
The use of ad-hoc engineered viruses in the fight against tumours is one of the greatest ideas in cancer therapeutics within the last three decades. Although some remarkable successes have been obtained, it is still not entirely clear how to achieve reliable protocols that can be routinely employed with confidence on a significant range of tumours. In this work, we concentrate on the study of different mathematical descriptions of virotherapy with the aim of better understanding the role of viral infectivity and viral dynamics in positive therapeutic outcomes. 
	
In particular, we compare probabilistic, individual approaches with continuous, spatially inhomogeneous models and investigate the importance of different tumour motility and different mathematical representations of viral infectivity. Some of these formulations also allow us to arrive at better analytical characterisation of how waves of viral infections arise and propagate in tumours, providing interesting insights into therapy dynamics.
		
Similarly to previous studies, oscillatory behaviours, stochasticity and cancers' diffusivities are all central to the eradication or the escape of tumours under virotherapy. Here, though, our results also show that the ability of viruses to  infect tumours seems, in certain cases, more important to a final positive outcome than tumours' motility or even reproducibility. This could hopefully represent a first step into better insights into viral dynamics that may help clinicians to achieve consistently better outcomes.
\end{abstract}
	

\section{Introduction}
	
	Oncolytic virotherapy is an anti-cancer technique that uses genetically modified viruses to directly infiltrate and destroy tumour cells, as well as to elicit an anti-cancer immune response \cite{blanchette23, fountzilas17, Kelly2007651review, lawler17,russell18}. Although it has been known for more than a century that external pathogens can provoke and sustain an immune reaction beneficial to cancer patients, systematic investigations and clinical trials on virotherapy as a viable cure against some tumours have only started a couple of decades ago. Notwithstanding some successes \cite{shalhout23}, there are still many unknown and complex factors that have not been completely understood and are preventing a full adoption of virotherapy in the clinics.
	Furthermore and as a result of the shortcomings for some stand-alone virotherapy approaches \cite{martin18},  combinations with a number of other therapeutics is also currently being attempted, including, for example, immunotherapies~\cite{engeland22}, immune checkpoint inhibitors~\cite{Sharma2023Checkpoint}, chimeric receptor T-cells~\cite{Sterner2021CarT} and others ~\cite{Mantovani2022Macro, Liu2024Other}. 
	
	There are many approaches in the mathematical literature that aim at describing the interaction between cancers, viruses and the immune systems. For example, ordinary differential equations (ODEs) have extensively been used~\cite{almuallem21, eftimie18, storey20, vithanage23, wodarz01}, as well as partial differential equations (PDEs) \cite{friedman18,kim18,lee20,wu04}, stochastic agent-based models (ABMs) \cite{storey21,surendran23} and hybrid discrete-continuous multi-scale models \cite{jenner22,paiva09}. The present work uses some of the insights from previous works, focusing on the role of an explicit formulation of the dynamics of viral infection. This aspect represents one of the sensitive points of the approach and its analysis can help improve the general understanding of virotherapy. 
	
	In our previous works \cite{morselli23,morselli24}, we have modelled infection of tumour cells due to oncolytic viruses without an explicit term for viral particles, i.e. the infection have been implicit and rate-based, since there have been no diffusive and/or motility terms for the viral vectors. We have also proposed two alternative sets of rules for cancer cell movement, based on unhindered or constrained cellular motility. In this work, we instead investigate the effect of an explicit viral term and its impact on outcomes for different movement rules for tumours, with the aim of understanding if and how eradication is governed by viral dynamics. In line with our previous approaches, we employ both an individual-based probabilistic model for single cells' and viral vectors' rules and a deterministic continuum model for volume fractions and concentrations. Standard methods are used to derive the continuum macroscopic model from the underlying discrete stochastic model (see, for example,  Refs. \cite{chaplain20, champagnat07, johnston15,lorenzimurray20, macfarlane22, penington11}). By comparing the different schemes, stochasticity, oscillations and tumour escape are investigated in the light of the mechanisms for viral infection; the specific focus on viral dynamics is the main novelty of this work. Furthermore, this comparison also allows us to refine the current knowledge on how waves of infections propagate, for small and large cell population numbers, and on how accurately the wave speed can be predicted using an analytic approach.
	
	To highlight why this perspective is important, let us remark that infections may take place via a number of mechanisms, including direct cell-to-cell transmission and cell-free transmission mediated by diffusing virions; the actual combination of processes is hard to establish in full detail (see Refs. \cite{graw16, Cohen2016virusInv} and the references therein). The two mechanisms appear relatively interchangeable in non-spatial models; on the other hand, spatial models may highlight some selected features of the two mechanisms \cite{williams25}. In the specific case of oncolytic virotherapy, there are also several obstacles to viral diffusion in the tumour microenvironment: indeed, factors such as the extracellular matrix composition, immune cell infiltration, tumour stroma and hypoxic regions can impede viral penetration, replication and spread within the tumour \cite{wojton10}. These considerations suggest that viral spread far from infected cells could play a minor role in virotherapy with respect to the infections caused by cell-to-cell contact, membrane fusion, endocytosis and close-range free virions. As a consequence, a common choice has been to neglect the complexities of viral dynamics and only focus on uninfected and infected cells: this approach has been used for non-spatial models of oncolytic viruses \cite{komarova10,novozhilov06} and a few spatial models \cite{morselli23,wodarz12}. On the other hand, the uncertainties highlighted above motivate our interest in exploring how the addition of explicit viral dynamics affects the outcome, at least at a theoretical level. We then model the decreased viral displacement by changing the diffusion coefficient, although direct modifications of the equation that directly incorporate obstacles in the diffusion are also possible \cite{pooladvand22}. Our results show that, in most cases, this is comparable to the two-equations model obtained by assuming quasi-stationary viral dynamics. On the other hand, a good viral infiltration of the tumour is associated with a treatment outcome that is at least partially favourable; the mathematical model then provides suggestions on how to improve the outcome.
	
	The paper is organised as follows. \rev{In Section \ref{sec:DCvirus},} we derive and discuss the models we use, highlighting in detail the hypotheses and the conditions that are typical of the ABMs, PDEs and ODEs approaches for virotherapy. \rev{Section \ref{sec:results}} contains original results on various typical scenarios, analysing the effects of stochasticity and viral infectivity on eradication, oscillations and tumour escape. Some new results on wave propagation dynamics, analytical evaluation of typical wave speeds and different tumour diffusivity are also of note. Finally, \rev{Section \ref{sec:conclusion}} critically discusses the findings and the limits of the different systems, suggests some insights with possible therapeutic benefits and provides  considerations on possible extensions and new avenues for research.

	\section{Model description}
	\label{sec:DCvirus}

	\subsection{Description of the agent-based models}
	
	We build upon the modelling framework presented in Ref. \cite{morselli23} and include viral particles, which were not modelled in our previous work: their concentration is described by a discrete non-negative function; as a consequence, the modelling framework is hybrid discrete-continuous. For the sake of simplicity, we restrict our attention to one spatial dimension in the formulation of the models and refer to \rev{Remark}~\ref{rmk:dif_twod} for the adaptation to two spatial dimensions.
	
	Let us consider the temporal discretisation $t_n=\tau n$ with $n\in\N_0$, $0<\tau \ll 1$ and the spatial discretisation $x_j=\delta j$, with $j\in\Z$, $0<\delta\ll 1$; we assume $\tau$ to be small enough to guarantee that all the probabilities defined hereafter are smaller than $1$. We denote the number of uninfected and infected cells that occupy position $x_j$ at time $t_n$ respectively by $U_j^n$ and $I_j^n$; the corresponding densities are
	\[
	u_j^n\coloneqq \frac{U_j^n}{\delta}, \qquad i_j^n\coloneqq \frac{I_j^n}{\delta}
	\]
	We denote by $\rho_j^n$ the local pressure at time $t_n$ and position $x_j$; we assume it to be equal to the local cell density, so that $\rho_j^n\coloneqq u_j^n+i_j^n$. In principle, we could also consider more general expressions for the pressure, such as the functional form proposed in Ref. \cite{perthame14}; while there might be differences in the way the system reaches the carrying capacity and how the model appears in the continuum limit, it has been shown that the overall behaviour of the tumour is not profoundly affected by different functional forms \cite{macfarlane22}.
	
	We then denote by $v_j^n$ the concentration of viral particles at time $t_n$ and position $x_j$. Since the spatio-temporal scales for the viruses' dynamics are very different from cellular ones, we describe them with a discrete balance equation; we remark that this equation contains stochastic elements, as will be explained shortly.
	
	The rules governing the dynamics of the agents correspond precisely to the ones presented in Ref. \cite{morselli23}, with the only exception of the infection, which here involves the viral concentration.  We consider both movement mechanisms described previously, i.e. undirected and pressure-driven, giving rise to different models. We then consider an additional equation for the evolution of the viral density.

	\paragraph{Dynamics of cancer cells}
	Proliferation, undirected movement and infection of uninfected cancer cells follow the same rules as in Ref. \cite{morselli23}, which we recall here briefly for the sake of completeness.
	
	We let an uninfected cell that occupies position $x_j$ at time $t_n$ reproduce with probability $\tau G(\rho_j^n)_+$, die with probability $\tau G(\rho_j^n)_-$, and remain quiescent with probability $1-\tau G(\rho_j^n)_+-\tau G(\rho_j^n)_-=1-\tau \abs{G(\rho_j^n)}$. We set
	\[
	G(\rho)=p\Bigl(1-\frac{\rho}{K}\,\Bigr)
	\]
	where $p>0$ is the maximal duplication rate and $K>0$ is the carrying capacity.
	
	We denote by $F_{j\to j\pm 1}^n$ the probability that a cancer cell that occupies position $x_j$ at time $t_n$ moves to the lattice point $x_{j\pm 1}$. We assume either $F_{j\to j\pm 1}^n\coloneqq \theta/2$, corresponding to undirected movement, or 
	\[
	F_{j\to j\pm 1}^n\coloneqq \theta \frac{(\rho_j^n-\rho_{j\pm 1}^n)_+}{2K}
	\]
	which corresponds to pressure-driven movement; in both cases $\theta\in[0,1]$. Cells may also remain at the initial position: this happens with probability $1-F_{j\to j- 1}^n-F_{j\to j+ 1}^n$.
	
	\rev{We also assume that at every time step an infected cell may die because of lysis with probability $\tau q$, where $q>0$ is a constant death rate.} The infection is here caused by contact between uninfected cells and viral particles. This means that an uninfected cell that occupies position $x_j$ at time $t_n$ becomes infected with probability $\tau\bthree v_j^n/K$, where $K$ is the carrying capacity and $\bthree>0$ is a constant infection rate. 
	Note the different approach with respect to our previous works, in which we assume that an uninfected cell becomes infected upon contact with another infected cell that occupies the same position \cite{morselli23,morselli24}. \rev{This} difference is the key ingredient to compare cell-to-cell infection with the infection mediated by free virions.

	\paragraph{Dynamics of the virus}
	We assume that infected cells release viral particles when they die due to lysis. Since cell death is stochastic, the balance equation for viral density needs to take it into account. The virus also decays at rate $q_v>0$ and diffuses. \rev{We use the notation $q_v$ so that parameters related to decays are labelled in a consistent way.}  Here, we do \rev{not} consider the uptake of free virions by cancer cells: a few models include this process \cite{bajzer08,pooladvand21}, but it is commonly assumed to be negligible. The resulting balance equation is finally:
	\begin{equation}
		\label{eq:v_discrete}
		v_j^{n+1}=v_j^n+\tau D_v \frac{v_{j+1}^n+v_{j-1}^n-2v_j^n}{\delta^2} +\frac{\alpha L_j^n}{\delta}-\tau q_v v_j^n
	\end{equation}
	where $D_v>0$ is the diffusion coefficient and $L_j^n$ denotes the number of infected cells at position $x_j$ that die at time $t_n$; this last term is multiplied by $\alpha$, which is the number of viral particles released, and divided by $\delta$ to obtain the density from the number of particles. This equation closely resembles the ones used in Refs. \cite{almeida22,bubba20,morselli24} to model the evolution of a chemoattractant concentration, with the main difference given by the presence of stochasticity.
	
	\subsection{Corresponding continuum models}
	
	Letting $\tau,\delta\to 0$ in such a way that \rev{$\theta\frac{\delta^2}{2\tau}\to D_k$ (with $k=U,P$ in the two different models), we arrive at the continuum counterparts of the individual models described above. The values of the diffusion coefficients are chosen to ensure identical propagation speeds in both models (see Table~\ref{tab:parametersB1} and Appendix~\ref{app:wave_u} for details).} 
	
	In the case of cancer cells, the derivation is almost identical to the calculations performed in Ref. \cite{morselli23}, with the only difference that now the term $\tau\bthree v_j^n/K$ replaces $\tau{\btwo} i_j^n/K$; for the sake of brevity, we do not report the details. On the other hand, the continuum counterpart of Eq. \eqref{eq:v_discrete} follows the same approach adopted for the chemoattractant in Refs. \cite{almeida22,bubba20,morselli24}, which we now sketch: we first take the expected values of Eq. \eqref{eq:v_discrete} and rearrange the terms to obtain
	\[
	\frac{v_j^{n+1}-v_j^n}{\tau}= D_v \frac{v_{j+1}^n+v_{j-1}^n-2v_j^n}{\delta^2} + \alpha \frac{q\mathbb{E}[I_j^n]}{\delta}- q_v v_j^n
	\]
	where $\mathbb{E}[\cdot]$ is the expected value.
	We then assume that there is a function $v\in C^2([0,+\infty)\times\R)$ such that $v_{j}^{n}=v(t_n,x_j)=v$ and let $\tau,\delta\to 0$ to obtain
	\[
	\pt v=D_v \pxx v +\alpha q i-q_v v
	\]

	In the case of undirected cell movement, we obtain the following system of reaction-diffusion PDEs:
	\begin{equation}
		\label{eq:lv}\tag{U3}
		\begin{cases}
			\pt u=\rev{D_U}\pxx u+pu\Bigl(1-\dfrac{u+i}{K}\Bigr)-\dfrac{\bthree}{K} uv \\[8pt]
			\pt i=\rev{D_U}\pxx i+\dfrac{\bthree}{K} uv-qi\\[8pt]
			\pt v=D_v \pxx v +\alpha q i-q_v v
		\end{cases}
	\end{equation}
	Note that this model has already been introduced in Ref. \cite{baabdulla23} (although the derivation from the discrete counterpart, to our knowledge, has not been previously performed) and it is a slightly different version of the approach in Ref. \cite{pooladvand21}.
	
	In the case of pressure-driven cell movement, we have the following local cross-diffusion system:
	\begin{equation}
		\label{eq:pv}\tag{P3}
		\begin{cases}
			\pt u=\dfrac{\rev{D_P}}{K} \px [u\px (u+i)]+pu\Bigl(1-\dfrac{u+i}{K}\Bigr)-\dfrac{\bthree}{K} uv \\[8pt]
			\pt i=\dfrac{\rev{D_P}}{K} \px [i\px (u+i)]+\dfrac{\bthree}{K} uv-qi \\[8pt]
			\pt v=D_v \pxx v +\alpha q i-q_v v
		\end{cases}
	\end{equation}
	This system shares some similarities with the ones presented in Refs. \cite{byrne09,gwiazda19,perthame14}, although we are not aware of other works that combine cross-diffusion and standard diffusion in the context of infections. \rev{The well-posedness of similar kinds of cross-diffusion ``triangular'' systems (i.e., with some equations only containing standard diffusion) have been previously studied \cite{brocchieri24,conforto18,desvillettes15,desvillettes23,yamada95}, but to our knowledge the existing results cannot be directly applied to Eq. \eqref{eq:pv}.}
	We will see shortly that viral spatial diffusion dampens several of the characteristic features described in the case of pressure-driven movement \rev{studied} previously \cite{morselli23}. It is also interesting to compare Eq. \eqref{eq:pv} with Eq. \eqref{eq:p} that will be introduced shortly to highlight the important differences in viral spreading and therapeutic efficiency related to different mechanisms of tumour cell's movement, especially when viral diffusion is partially \rev{constrained}.

	\begin{rmrk}
	    \label{rmk:dif_twod}
		When the spatial domain is the two-dimensional real plane $\R^2$ instead of the one-dimensional real line $\R$, the scalar index $j\in\Z$ should be replaced by the vector $\vec{j}=(j_x,j_y)\in\Z^2$ and the probability that a cell moves to one of the four neighbouring lattice points is $\theta_k/4$ in the case of undirected movement and
		\[
		\theta_k \frac{(\rho_{\vec{j}}^n-\rho_{\vec{j}+\vec{e}}^n)_+}{4K}
		\]
		with $k=u,i$ and $\vec{e}\in \{(\pm 1,0),(0,\pm 1)\}$ in the case of pressure-driven movement. We then need to scale $\tau$ and $\delta$ in such a way that \rev{$\frac{\delta^2}{4\tau}\to D_k$, with $k=U,P$ in the two different models}. 
	\end{rmrk}

	\subsection{Quasi-steady viral dynamics in non-spatial models}
	
	In view of the following analyses, it is useful to consider the non-spatial version of Eqs. \eqref{eq:lv} and \eqref{eq:pv} to discuss the typical behaviour of equilibria for homogeneous ODE formulations:
	\begin{equation}
		\label{eq:odev}
		\begin{cases}
			\dfrac{\di u}{\di t}=pu\Bigl(1-\dfrac{u+i}{K}\Bigr)-\dfrac{\bthree}{K} uv \\[8pt]
			\dfrac{\di i}{\di t}=\dfrac{\bthree}{K} uv-qi \\[8pt]
			\dfrac{\di v}{\di t}=\alpha q i-q_v v
		\end{cases}
	\end{equation}
	
	Viral dynamics are much faster than cells' dynamics: as a consequence, non-spatial models of oncolytic viruses such as the one in Eq. \eqref{eq:odev} often rely on the quasi-steady assumption $\frac{\di v}{\di t}\approx 0$, which implies that
	\begin{equation}
		\label{eq:vsteady}
		v\approx\frac{\alpha q}{q_v} i
	\end{equation}
	We can then define the cell-to-cell infection rate
	\begin{equation}
		\label{eq:betatilde}
		\btwo\coloneqq\frac{\bthree \alpha q }{q_v}
	\end{equation}
	and reduce Eq. \eqref{eq:odev} to the system
	\begin{equation}
		\label{eq:ode}
		\begin{cases}
			\dfrac{\di u}{\di t}=pu\Bigl(1-\dfrac{u+i}{K}\Bigr)-\dfrac{\btwo}{K} ui \\[8pt]
			\dfrac{\di i}{\di t}=\dfrac{\btwo}{K} ui-qi
		\end{cases}
	\end{equation}
	This choice allows to condense into a single parameter $\btwo$ the uncertainties of several biological processes related to the virus, which are not entirely known, and simplify the mathematical analysis (see, for example, Refs. \cite{komarova10,novozhilov06}). Nevertheless, it is important to remark that this system does not exhibit the same oscillatory behaviours of Eq. \eqref{eq:odev} and this may lead to overlook some relevant situations, as we discuss in Subsection \ref{sec:oscillations}.
	
	As it is well-known, Eq. \eqref{eq:ode} has three equilibria: $(0,0)$, $(K,0)$ and 
	\begin{equation}
		\label{eq:eq_l}
		(u^*,i^*)\coloneqq \left( \frac{qK}{\btwo}, \frac{pK(\btwo -q)}{\btwo(\btwo +p)}\right)
	\end{equation}
	The first equilibrium point is always unstable. The second one is stable when $\btwo<q$ (i.e., $i^*<0$) and unstable when $\btwo>q$ (i.e., $i^*>0$); in this last case, the third equilibrium is stable (see also Ref. \cite{morselli23}). 
	
	Eq. \eqref{eq:odev} has the same three equilibria, with $\btwo$ as in Eq. \eqref{eq:betatilde} and the viral concentration given by Eq. \eqref{eq:vsteady}. The first two equilibria have the same stability as in the simplified model. The third one is
	\begin{equation}
		\label{eq:eq_lv}
		\Bigl( u^*, i^*, \frac{\alpha q}{q_v}\, i^*\Bigr)=\left( \frac{qK}{\btwo}, \frac{pK(\btwo -q)}{\btwo(\btwo +p)},\frac{\alpha q}{q_v} \frac{pK(\btwo -q)}{\btwo(\btwo +p)}\right)
	\end{equation}
	and it may become unstable for $\btwo>q$ when a stable limit cycle arises (as described in Refs. \cite{baabdulla23,bansod26}). Let us remark that the spatially homogeneous steady-state of the tumour depends on $\alpha$ and \rev{$q_v$} only through $\btwo$, hence we should expect \rev{similar dynamics for parameter combinations that maintain the ratio $\alpha/q_v$ unchanged.} This situation is analysed in Subsection~\ref{sec:waves_sim}.

	\subsection{Quasi-steady viral dynamics in spatial models}
	
	In the context of spatial models, the different time scale of the processes alone would not justify neglecting viral diffusion: indeed, we will see in the following that significantly different behaviours may arise. Nevertheless, in some biological scenarios it is reasonable to assume that viral particles do not spread far from infected cells, as explained in the Introduction. 
	When this is the case, the infection is mainly driven by cell-to-cell contact and close-range free virions, hence \rev{it} is acceptable to ignore viral dynamics. All this leads to the equation
	\begin{equation}
		\label{eq:l}\tag{U2}
		\begin{cases}
			\pt u=\rev{D_U}\pxx u+pu\Bigl(1-\dfrac{u+i}{K}\Bigr)-\dfrac{\btwo}{K} ui \\[8pt]
			\pt i=\rev{D_U}\pxx i+\dfrac{\btwo}{K} ui-qi
		\end{cases}
	\end{equation}
	in the case of undirected movement (analogous to Eq. \eqref{eq:lv}) and 
	\begin{equation}
		\label{eq:p}\tag{P2}
		\begin{cases}
			\pt u=\dfrac{\rev{D_P}}{K} \px [u\px (u+i)]+pu\Bigl(1-\dfrac{u+i}{K}\Bigr)-\dfrac{\btwo}{K} ui \\[8pt]
			\pt i=\dfrac{\rev{D_P}}{K} \px [i\px (u+i)]+\dfrac{\btwo}{K} ui-qi
		\end{cases}
	\end{equation}
	in the case of pressure-driven movement (analogous to Eq. \eqref{eq:pv}). \rev{The quasi-steady approximation of the virus can be mathematically recovered in the limit of fast viral dynamics and slow diffusion. The results of Section \ref{sec:lowdiff} indeed confirm the validity of the assumption in this parameter regime.}
	
	Both these models have been studied previously \cite{morselli23}. Eq. \eqref{eq:l} fall in the category of classical spatial Lotka--Volterra models for prey and predators, which have been widely studied \cite{dunbar84,li15,morozov06,petrovskii05,petrovskii02}. Eq. \eqref{eq:p} replaces the standard diffusion with nonlinear cross-diffusion: similar models have been introduced in several other works \cite{bubbaheleshaw20,carrillo18,gwiazda19,lorenzi17}, but to our knowledge the exact formulation of Eq. \eqref{eq:p} has only been presented in our previous work \cite{morselli23}.

	\section{Results}
	\label{sec:results}
	
\begin{table}[t!]
		\centering{
        \scriptsize
			\begin{tabular}{lllll}
				\toprule
				&\multicolumn{1}{c}{\textbf{Parameter}} & \multicolumn{1}{c}{\textbf{Description}} &  \multicolumn{1}{c}{\textbf{Value [Units]}} & \multicolumn{1}{c}{\textbf{Reference}}\\
				\midrule
				& $p$         &maximal duplication rate of uninfected cells     & $1.87 \times 10^{-2}$ [h$^{-1}$] &\cite{ke00} \\
				& $q$         &death rate of infected cells     & $4.17\times 10^{-2}$ [h$^{-1}$]	&\cite{ganly00} \\
				& $\rev{D_U}$         &diffusion coefficients (undirected movement)    & $1.88\times 10^{-4}$ [mm$^2$/h]	&estimate based on \cite{kim06} \\
				& $\rev{D_P}$         &diffusion coefficients (pressure-driven movement)    & $1.50\times 10^{-3}$ [mm$^2$/h]	&estimate based on \cite{kim06} \\
				& $K^{1 \text{D}}$         &tissue carrying capacity in one dimension   & $10^3$ [cells/mm]	&model estimate \\
				& $K^{2 \text{D}}$         &tissue carrying capacity in two dimensions   & $10^4$ [cells/mm$^2$]	&\cite{lodish08} \\
				\midrule
				& $q_v$ 			&virus clearance rate		&$1.67\times 10^{-1}$ or $1.00$ [h$^{-1}$] &\cite{mok09}, model estimate \\
				& $\alpha$         &viral burst size     & $580$ or $3500$ [viruses/cells]	&\cite{workenhe14,chen01}\\
				& $\bthree$         &infection rate   & $7.00\times 10^{-4}$ [viruses/(cells$\cdot$h)]	&\cite{friedman06} \\
				\midrule
				& $D_v$         &virus diffusion coefficient    & $1.00\times 10^{-2}$ [mm$^2$/h]	&\cite{pooladvand21} \\
				& $R_u$         &initial radius of uninfected cells     & $2.60$ [mm]	&\cite{kim06} \\
				& $R_v$         &initial radius of oncolytic virus     & $0.50$ [mm]	&\cite{pooladvand21} \\
				& $R_i$         &initial radius of infected cells     & $0$ [mm]	&model estimate \\
				\bottomrule
			\end{tabular}
			\caption{Reference parameter values. \rev{Note that the value of $\btwo$ is given by Eq. \eqref{eq:betatilde} and depends on $\bthree, \alpha$ and $q_v$.}}\label{tab:parametersB1}}
\end{table}	
	
	We now compare numerical simulations for the agent-based models and the corresponding PDE systems. 
	We begin by describing the details of numerical simulations (Subsection \ref{sec:num}). Next, we examine the emergence of travelling waves and some of their quantitative features (Subsection \ref{sec:waves_sim}). We then turn to the case of low viral diffusion, where different models display significant discrepancies (Subsection \ref{sec:lowdiff}). Finally, we describe the appearance of oscillations, which do not occur for a quasi-steady virus, and discuss the possibility of tumour eradication (Subsection \ref{sec:oscillations}).
	
	\subsection{Details of numerical simulations}
	\label{sec:num}
	We use the following initial conditions:
	\begin{equation}
		\label{eq:initialv}
		u_0(x)=
		\begin{cases}
			0.9\ K \quad &\text{for } \abs{x}\leq R_u \\
			0\ \quad &\text{for } \abs{x}> R_u 
		\end{cases}
		\qquad
		i_0(x)=0
		\qquad
		v_0(x)=
		\begin{cases}
			V_0 \quad &\text{for } \abs{x}\leq R_v \\
			0 \quad &\text{for } \abs{x}> R_v
		\end{cases}
	\end{equation}
	with $V_0=2.67\times 10^4$ in one spatial dimension and $V_0=(2.67\times 10^3)^2$ in two spatial dimensions; these values are obtained by appropriately rescaling the one used in Ref. \cite{pooladvand21}, which comes from the number of viral particles used in Ref. \cite{kim06}. The same initial conditions are also employed for the agent-based models.
	
	The parameters are set to the same values adopted in Ref. \cite{morselli23}, with the only exception of \rev{the infection rate $\bthree$, the virus clearance rate $q_v$, the viral burst size $\alpha$, the viral diffusion coefficient $D_v$ and the initial radius of oncolytic virus $R_v$, which} are not present in those models. We briefly explain our choice of the values and report the full parameters list in Table~\ref{tab:parametersB1} for the reader's convenience. The spatial diffusion coefficient of viral particles $D_v$ has been set to $1.00\times 10^{-2}\;$mm$^2$/h, as in Ref. \cite{paiva09}; this value is in line with other estimates found in literature \cite{friedman06}. The viral injection radius $R_v$ is taken equal to $0.5\;$mm, following Ref. \cite{pooladvand21}. 
	The infection rate $\bthree$ has been estimated from Ref. \cite{friedman06}: there, the authors assume that oncolytic viruses have an infection rate in three spatial dimensions of $7\times 10^{-10}\;$mm$^3$/(viruses$\times$h) and cells have a carrying capacity of $10^6\;$cells/mm$^3$; in our model the infection rate is always divided by the carrying capacity to simplify the scaling in different dimensions, hence the value of $\bthree$ has been rescaled appropriately. 
	
	The viral load released by the death of infected cells $\alpha$ depends highly on the type of virus and ranges from the value $157\;$viruses/cells estimated in Ref. \cite{workenhe14} to the value $3500\;$viruses/cells of Ref. \cite{chen01}; this wide variation is among the motivations that led us to consider different values of $\alpha$. The value of Ref. \cite{workenhe14} however appears too small to show an effective infection in our settings, hence in our simulation we use a minimum value of $\alpha$ equal to $580\;$viruses/cells. The clearance rate of the virus is also a parameter that could significantly vary in different situations: in Ref. \cite{mok09}, it has been estimated to $1.67\times 10^{-1}\;$h$^{-1}$ \textit{in vitro}, but it is reasonable to assume it to be higher in different settings, due for example to the immune response \textit{in vivo} or other factors that reduce the viral lifespan. We therefore also consider a higher value of $1.00\;$h$^{-1}$. We finally remark that an adequate comparison requires the adaptation of the diffusion coefficient to keep the propagation speed constant in the different models (as listed in Table~\ref{tab:parametersB1} and further explained in Appendix \ref{app:wave_u}).
	
	All the simulations have been performed in \textsc{Matlab 2021b} with the techniques described in Ref. \cite{morselli23}. 
	For the individual-based models, we used a temporal step $\tau=0.02\;$h and a spatial step of $\delta=0.1\;$mm both for the one-dimensional and the two-dimensional simulations. Since we only need to keep track of the collective fate of cells in the same lattice point, we used the built-in \textsc{Matlab} functions \texttt{binornd} and \texttt{mnrnd}, which compute random arrays according to binomial and multinomial distributions. In order to allow reproducibility, a random seed has been set at the beginning of each new simulation, ranging from one to five (except in the case of Fig. \ref{fig:violin}, which shows the results of one hundred simulations). In the figures representing a single simulation, only the one with random seed equal to one is shown: we remark that we performed at least five different simulations also in these cases and the resulting stochastic behaviour arises in all of them in a qualitatively similar way (although the exact quantitative features are clearly different).
	
	Eqs. \eqref{eq:lv} and \eqref{eq:l} have been solved with a finite difference scheme explicit in time, using the discretisations $\Delta x=0.1$, $\Delta t=10^{-3}$. We used a forward upwind scheme for the nonlinear diffusion of Eqs. \eqref{eq:pv} and \eqref{eq:p}, following Ref. \cite{leveque07}; the discretisations are again $\Delta x=0.1$, $\Delta t=10^{-3}$. We remark that the simulations of the electronic supplementary material S4 require a refined discretisation $\Delta x=0.01$, $\Delta t=10^{-4}$ to reduce the diffusion due to the numerical method. In the two-dimensional case we exploited the radial symmetry of the equations.
	
	\subsection{Partial treatment success: travelling waves}
	\label{sec:waves_sim}

	Let us first focus on the dynamics that do not present persistent, wide oscillations. The spatially homogeneous steady-state of the tumour is independent of $D_v$ and depends on $\alpha$ and $q$ only through $\btwo$. This observation is already enough to provide useful insights on how to improve the overall treatment outcome, which remain valid whenever the infection spreads in the whole tumour. Indeed, increasing $\btwo$ in any way and decreasing $q$ appears beneficial to the treatment. Even for parameter sets that maintain the equilibria unchanged, it is reasonable to expect differences in the spatial viral spread, which could significantly affect the treatment.

	\paragraph{Analytical results}
	The speed of travelling waves solutions for PDEs with spatial diffusion can be computed using linearisation techniques. In the case of Eq. \eqref{eq:l}, infected cells invade regions in which uninfected cells are at carrying capacity at speed \rev{$2\sqrt{D_U(\btwo-q)}$} and slow down as the uninfected cell density decreases \cite{morselli23}; the same result can be also obtained in a rigorous way for a similar system of PDEs \cite{dunbar84}. In the case of Eq. \eqref{eq:lv}, the minimal speed can be found at the intersection of two appropriate manifolds \cite{baabdulla23}: although it is hard to derive an explicit expression, we can easily rely on numerical computations (see Appendix~\ref{app:wave}). Fig. \ref{fig:speedqv} shows the behaviour of the minimal speed for different parameter values. For the sake of simplicity, we \rev{here} restrict our attention to parameter combinations that maintain the ratio $\alpha/q_v$ constant: while this is not a biological constraint, it allows to focus our attention to the specific differences between the modelling approaches in comparable settings. \rev{Other parameter ranges are explored in Fig. \ref{fig:speed_extra}.}
	
	\begin{figure}
		\centering
		\includegraphics[width=\linewidth]{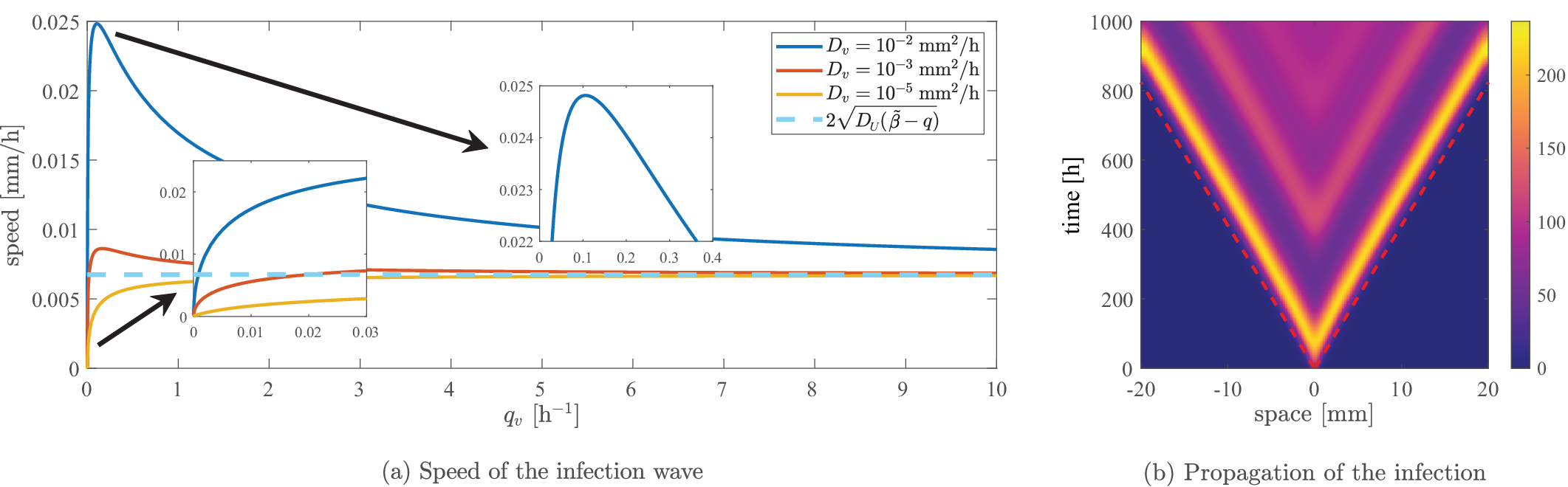}
		\caption{(a) Theoretical speed of the travelling wave of viral infection invading an uninfected tumour at carrying capacity, computed following the procedure of Ref. \cite{baabdulla23} (as explained in the Appendix \ref{app:wave}). The parameters employed are the ones given in Table~\ref{tab:parametersB1}, with three values of $D_v$; $q_v$ and $\alpha$ are varied so that the value of $\btwo$ is constantly equal to $1.02\times 10^{-1}\;$h$^{-1}$. The dashed cyan line shows the speed of the infection wave obtained for Eq. \eqref{eq:l} with the same parameters. (b) Numerical solutions of Eq. \eqref{eq:lv} compared with the position of the theoretical front (shown as a dashed red line). The parameters employed are the ones given in Table~\ref{tab:parametersB1}, with $\alpha=580\;$viruses/cells and $q_v=1.67\times 10^{-1}\;$h$^{-1}$; we also set \rev{$u(0,\cdot)=K$} to focus on the infection inside an established tumour.}
		\label{fig:speedqv}
	\end{figure}

	\rev{Fig. \ref{fig:speedqv} shows that, as} $q_v$ grows, the speed of the infection wave converges to the value \rev{$2\sqrt{D_U(\btwo-q)}$}, meaning that viral dynamics are irrelevant because the virus is cleared very quickly. Instead, the case $q_v=0$ corresponds to $\alpha=0$, which means that viral particles only diffuse without any production or decay; travelling waves are not possible in this hypothetical situation, even though this state has no biological meaning. It is also interesting to observe that for very low values of $q_v$ the speed is extremely low: in this case, viral dynamics are mostly dictated by spatial diffusion and the effects of the reaction part of the equation is negligible. In the intermediate cases, the speed is either higher or lower than \rev{$2\sqrt{D_U(\btwo-q)}$} depending on whether $D_v$ is higher or lower than $\rev{D_U}$ \rev{(see also Fig. \ref{fig:speed_extra}b)}. It is \rev{interesting to observe that when $D_U$ is high enough there is a specific value of $q_v$ that maximises the speed among the parameter sets that maintain the ratio $\alpha/q_v$ constant.}
	
	To our knowledge, travelling wave solutions for PDEs of the kind of Eqs. \eqref{eq:pv} and \eqref{eq:p} have not been characterised in the cases of our interest: the techniques previously mentioned do not provide any meaningful information on such intrinsically nonlinear equations.

	\paragraph{Undirected movement}
	
	\begin{figure}
		\centering
		\includegraphics[width=\linewidth]{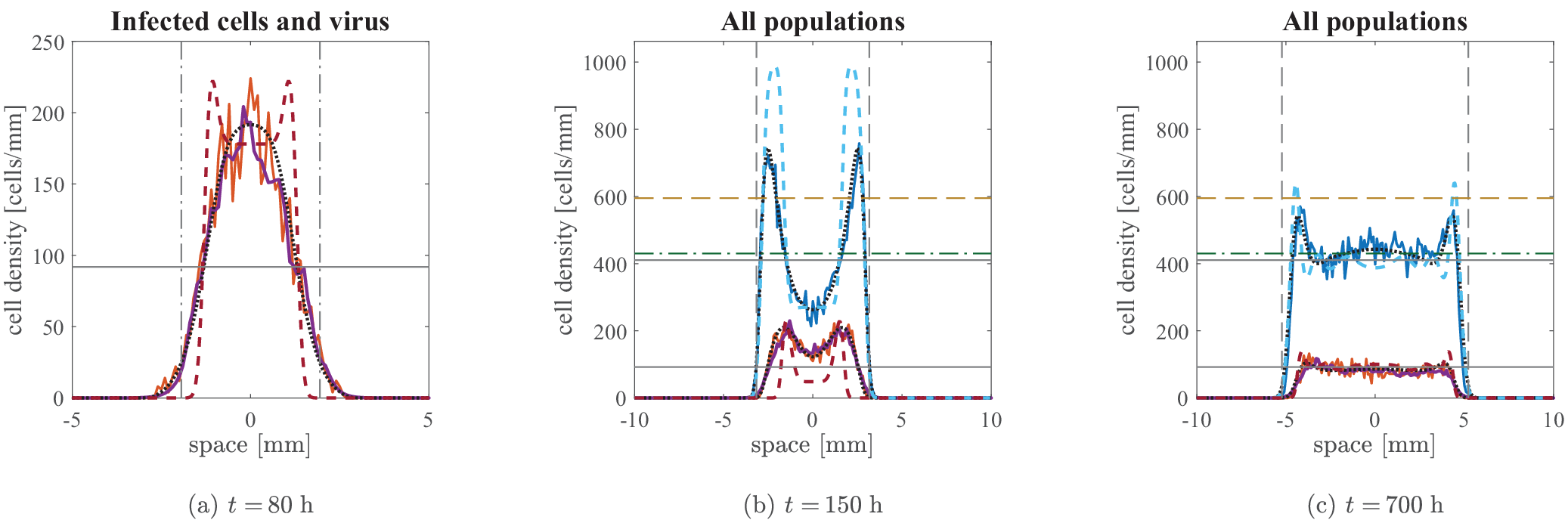}
		\caption[One-dimensional simulations with viral dynamics, undirected cell movement, low viral burst size and low viral decay]{Comparison in one spatial dimension between numerical simulations of the discrete model with undirected cell movement (solid lines), the numerical solution of Eq. \eqref{eq:l} (dashed lines) and the numerical solution of Eq. \eqref{eq:lv} (dotted black lines) at three different times, with the parameters given in Table \ref{tab:parametersB1}, $\alpha=580\;$viruses/cells and $q_v=1.67\times 10^{-1}\;$h$^{-1}$. The viral density of the agent-based model is multiplied by $q_v/(\alpha q)$ to allow the comparison with cell numbers; note that the viral density for Eq. \eqref{eq:lv} is not shown, as it would superimpose with infected cells. For the agent-based model, the densities of uninfected cells, infected cells and virus are represented respectively in blue, red and purple; the numerical solutions of Eq. \eqref{eq:l} are represented using the same colours. The vertical black lines represent the expected positions of the invasion fronts:  the dash-dotted line in panel (a) refers to the infected front, whose speed is \rev{depicted} in Fig.~\ref{fig:speedqv}, and the dashed line in panels (b) and (c) refers to the uninfected front, travelling at speed $2\sqrt{\rev{D_U}p}$. The horizontal solid black lines show the equilibrium of the ODE given by Eq. \eqref{eq:eq_lv}. The horizontal dashed yellow line represents the expected uninfected density at the front for Eq. \eqref{eq:l} given by Eq. \eqref{eq:ubar} and the horizontal dash-dotted green line shows the analogue quantity in the case of Eq. \eqref{eq:lv} (both are only relevant in panel (c)). The results of the agent-based model are averaged over five simulations. The maximum of the cell density axes in panels (b) and (c) corresponds to the maximum over time of this average (which is larger than the carrying capacity); on the other hand, in panel (a), it has been reduced to allow the comparison between infected cells and virus (and, as a consequence, uninfected cells are not shown).}
		\label{fig:lv_low}
	\end{figure}

	Let us now describe how the different speed of the infection affects the treatment in comparison to the previous approaches. We start by considering the lower values of $\alpha$ and $q_v$ from Table \ref{tab:parametersB1}, as in this case we expect a more significant difference. Fig.~\ref{fig:lv_low}, along with the video accompanying it (see electronic supplementary material S2), shows an excellent quantitative agreement between numerical solutions of the system of PDEs~\eqref{eq:lv} and the average over five numerical simulations of the corresponding agent-based model in one spatial dimension. Furthermore, the quasi-steady approximation of Eq. \eqref{eq:vsteady} holds most of the time, even in the agent-based model (despite some very small stochastic fluctuations), as it is evident from Fig. \ref{fig:lv_low}a. Nevertheless, the system's transient behaviour significantly differs from the solution of Eq. \eqref{eq:l}. Fig. \ref{fig:lv_low}a shows that the low viral decay allows the viral infection to reach the boundary of the tumour faster than in the case of cell-to-cell infection, in line with the higher speed predicted by theoretical results (see the vertical dash-dotted lines); as a consequence, uninfected cells start to decrease their numbers earlier (Fig. \ref{fig:lv_low}b). This difference does not significantly affect the system's overall behaviour: in all the cases, the travelling wave of uninfected cells invades the surrounding area at the speed $2\sqrt{\rev{D_U} p}$ (vertical dashed lines in Fig. \ref{fig:lv_low}b-c) and cell densities at the centre of the tumour converge with damped oscillations to the equilibrium of the corresponding ODE (horizontal solid black lines in Fig. \ref{fig:lv_low}), given by Eq. \eqref{eq:eq_l}.
	
	The final peak of the uninfected cells is approximately the one that allows the infection to travel at speed $2\sqrt{\rev{D_U}p}$: in the case of Eq. \eqref{eq:eq_l}, this is given by
	\begin{equation}
		\label{eq:ubar}
		\bar{u}\coloneqq \Bigl( \frac{q}{\beta}+\frac{p}{\beta} \Bigr) K= u^* +\frac{p K}{\beta}
	\end{equation}
	and it is shown as a yellow dashed line in Fig. \ref{fig:lv_low}b-c. In the full system, we observe a lower cell density at the front, which can be explained by the higher infiltration of the infection: the theoretical approximation (shown as a green dash-dotted line in Fig. \ref{fig:lv_low}b-c) predicts this decrease, but significantly underestimates the value; this disagreement is probably caused by the fact that the infection is not invading a spatially homogeneous uninfected population, which is the situation considered in the Appendix. Indeed, such a wave would have a longer spatial decay towards infinity, but this is not possible in the case of Fig. \ref{fig:lv_low}c because the uninfected population is more localised; as a consequence, the desired speed can only be obtained if the uninfected peak is slightly higher. In other words, nonlinear effects here play a more important role than in other settings.

	Let us now consider higher values of $\alpha$ and $q_v$, keeping their ratio approximately as in the previous case. The electronic supplementary material S3 shows again an excellent quantitative agreement between numerical solutions of the system of PDEs~\eqref{eq:lv} and the average over five numerical simulations of the corresponding agent-based model in one spatial dimension. Furthermore, the initial invasion of infected cells shows a larger degree of similarity to the one obtained by the two-equation model in Eq. \eqref{eq:l} than in the previous case, as expected by the value of the invasion speed. After an initial transient time, the uninfected cell density is almost indistinguishable from the one resulting from Eq. \eqref{eq:l}. Travelling waves are remarkably similar to the solution of Eq. \eqref{eq:l}, with almost the same density of uninfected cells at the front: the theoretical value is a much better approximation than in the previous case and correctly predicts the front to be slightly lower than for the two equations model.

	\paragraph{Pressure-driven movement}
	
	\begin{figure}
	\centering
	\includegraphics[width=\linewidth]{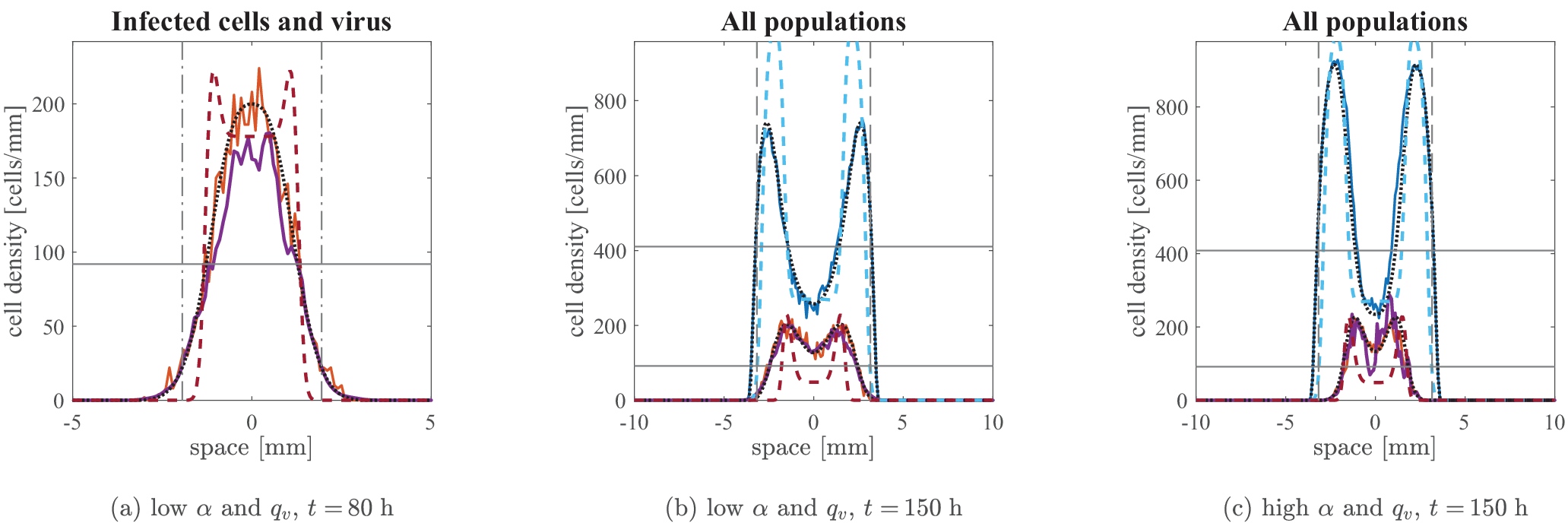}
	\caption[One-dimensional simulations with viral dynamics, pressure-driven cell movement, high viral burst size and high viral decay]{Comparison in one spatial dimension between numerical simulations of the discrete model with pressure-driven cell movement (solid lines), the numerical solution of Eq. \eqref{eq:l} (dashed lines) and the numerical solution of Eq. \eqref{eq:pv} (dotted black lines) at three different times, with the parameters given in Table \ref{tab:parametersB1} and both combinations of $\alpha$ and $q_v$. We remark that the diffusion coefficient in Eq.  \eqref{eq:l} was modified to keep the propagation speed constant (see Appendix \ref{app:wave_u}). All the graphical elements have the same meaning as in Fig. \ref{fig:lv_low}. Observe that, this time, the maximum value reached by the average uninfected cell density is lower than the carrying capacity.}
	\label{fig:pv}
\end{figure}

	Let us first consider the lower values of $\alpha$ and $q_v$. The behaviour of numerical simulations is very similar to the one observed in the previous case. Let us recall that, in the case of Eq. \eqref{eq:p}, central infections remain localised in the initial position, as described in Ref. \cite{morselli23} (see also Subsection \ref{sec:lowdiff}); this is in sharp contrast with the outcome shown in Fig. \ref{fig:pv}. Indeed, in the present situation, viral particles can reach the outer region of the tumour without any obstacles and the constraints of cell movement do not stop the propagation of the infection. The similarity of this situation to the case of undirected cell movement suggests that it is more appropriate to compare it to Eq. \eqref{eq:l} than to Eq. \eqref{eq:p}. Fig. \ref{fig:pv}a, along with the video accompanying it (see electronic supplementary material S2), shows that the spread of the infection is extremely similar with respect to the previous case, meaning that the viral diffusion affects the dynamics more significantly than cellular movement; small differences can only be observed by a close comparison of the PDE models, which however are totally unnoticeable in the agent-based model. The only relevant differences between undirected movement and standard diffusion appear to be the maximum value reached by uninfected cell density, which is higher in the former situation, and the profile of the final invasion front, which, in the latter case, is compactly supported and propagates to the surrounding tissues slightly faster. Overall, we can conclude that spatial viral dynamics have a more substantial effect on the evolution of the system than cellular spatial dynamics.
	
	Fig. \ref{fig:pv}c, along with the video accompanying it (see electronic supplementary material S3), shows that also for high $\alpha$ and $q_v$ the outcome is very similar to the case of undirected cell movement, which confirms our previous observation that the infective dynamics are affected mainly by viral spatial spread. This is not trivial\rev{.}

	\subsection{Possible treatment failure: low viral diffusion}
	\label{sec:lowdiff}
	
	In the previous situation, viral particles can reach the outer region of the tumour without any obstacles and the constraints of cell movement do not stop the propagation of the infection. We now analyse the behaviours that arise for small $D_v$ and focus on the influence of cell displacement to the spread of the infection. \rev{We remark that the case of fast viral dynamics and slow diffusion is the one in which we expect the quasi-steady assumption to be extremely accurate; as a consequence, we also expect Eqs.~\eqref{eq:l} and \eqref{eq:p} to approximate well Eqs.~\eqref{eq:lv} and \eqref{eq:pv}.}
	
	\begin{figure}
		\centering
		\includegraphics[width=\linewidth]{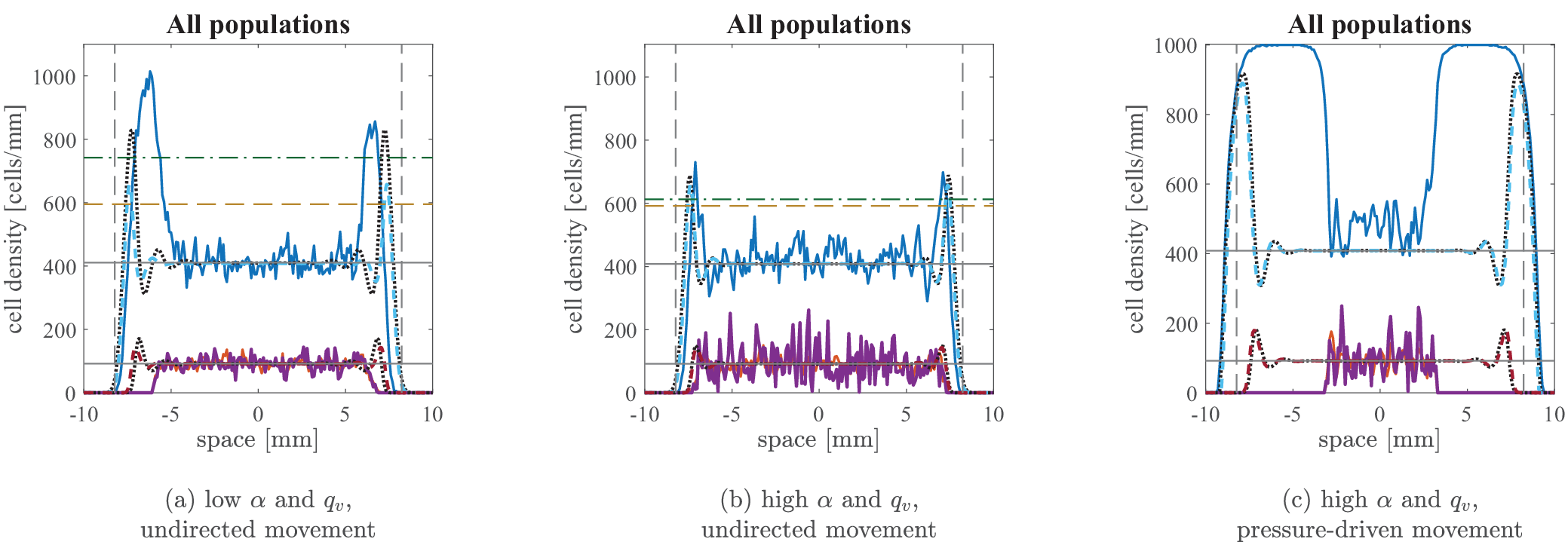}
		\caption{Comparison in one spatial dimension between the various models with a low value of viral diffusion $D_v=10^{-5}\;$mm$^2$/h at time $t=1500\;$h. Panels (a-b) show numerical simulations of the discrete model with undirected cell movement (solid lines), the numerical solution of Eq. \eqref{eq:l} (dashed lines) and the numerical solution of Eq. \eqref{eq:lv} (dotted black lines) with the parameters given in Table \ref{tab:parametersB1}, low viral diffusion and both combinations of $\alpha$ and $q_v$. 	Panel (c) shows numerical simulations of the discrete model with pressure-driven cell movement (solid lines), the numerical solution of Eq. \eqref{eq:p} (dashed lines) and the numerical solution of Eq. \eqref{eq:pv} (dotted black lines) with the parameters given in Table \ref{tab:parametersB1}, low viral diffusion, $\alpha=3500\;$viruses/cells and $q_v=1\;$h$^{-1}$; we also consider $R_v=R_u$ in the initial conditions to model a wide initial infection. All the graphical elements have the same meaning as in Fig. \ref{fig:lv_low}.}
		\label{fig:Dvlow}
	\end{figure}
	
	\paragraph{Undirected cell movement}
	Lower viral diffusion clearly implies a slower invasion speed of infection, as also shown in Figs. \ref{fig:speedqv} \rev{and \ref{fig:speed_extra}b}. The movement of cells is still sufficient for spreading the infection all over the tumour, as shown in Fig.~\ref{fig:Dvlow}a-b. It is also interesting to observe that lower values of $\alpha$ and $q_v$ are now associated with a slower speed, differently from the case of faster viral diffusion: indeed, viral particles with a shorter lifespan cannot effectively propagate for long distances.  Fig.~\ref{fig:Dvlow}a also shows that some stochastic events slow down the infection to the left of the domain, even though the results are averaged over five simulations; it appears therefore that stochasticity becomes more relevant for this less effective infection, although the overall behaviour is not affected significantly.  Fig.~\ref{fig:Dvlow}b shows that for higher values of $\alpha$ and $q_v$ the behaviour is almost indistinguishable to the one of Eq. \eqref{eq:l}: this is indeed the parameter range in which the quasi-steady assumption of the virus is most accurate. In both cases, the theoretical predictions of the peak of uninfected cells is very precise for the PDE and still constitutes a good approximation for the agent-based model.

	\paragraph{Pressure-driven cell movement}
	As the spatial spread of viral particles becomes less relevant, we expect to recover the main features of Eq. \eqref{eq:p}. First of all, a central infection remains mostly confined at the centre of the tumour for both parameter sets that we consider (only shown in the electronic supplementary material S4): the central infection causes the total cell density to drop, while external uninfected cells proliferate; since cells cannot move towards an area with higher cell density and viral diffusion is negligible, the outer cells are never going to be infected and the tumour keeps expanding in the same way it would do in the absence of treatment, almost identically to the situation described in Ref. \cite{morselli23}.
	
	We therefore focus on initial infections spread in the whole tumour, namely $R_v=R_u$, which gives rise to the formation of travelling waves for Eq. \eqref{eq:p} \cite{morselli23}. For the sake of brevity, we restrict our attention to higher values of $\alpha$ and $q_v$, which facilitate the spread of the infection for the continuous model; we remark that both cases are shown in electronic supplementary material S5. Fig. \ref{fig:Dvlow}c shows that the same kind of travelling waves is also obtained as a solution of Eq. \eqref{eq:pv} and the two solutions superimpose almost exactly; the peak of uninfected cells is higher than the one of the solutions of Eqs. \eqref{eq:l} and \eqref{eq:lv}, in line with the less effective spread of the infection. Note that the agent-based model again shows that the infection is confined to the centre of the tumour, in line with the case that neglects explicit viral dynamics. It is important to stress that, for the current model, demographic stochasticity plays a much more important role than in previous models: any growth above the average of uninfected cells stops the movement of infected cells and hence cannot be compensated at later times by other processes. 
	
	Taking the average over a higher number of realisations does not significantly improve the agreement, since the individual stochasticity of each simulation is not affected (not shown). According to the formal derivation of the PDEs from the agent-based model, an increase of cell number, ensuring that the evolution of the stochastic system diverges less significantly from the average behaviour described by the continuum model, and a decrease of the temporal and spatial discretisation all together improve the quality of the continuum approximation. In this specific case, the PDE predicts the presence of a very small infected cell density up to the uninfected invasion front; a good agreement would therefore require a number of infected cells high enough to neglect stochasticity. However, such an agreement can only be obtained at scales that are much larger than any meaningful biological process.

	\begin{figure}
		\centering
		\includegraphics[width=\linewidth]{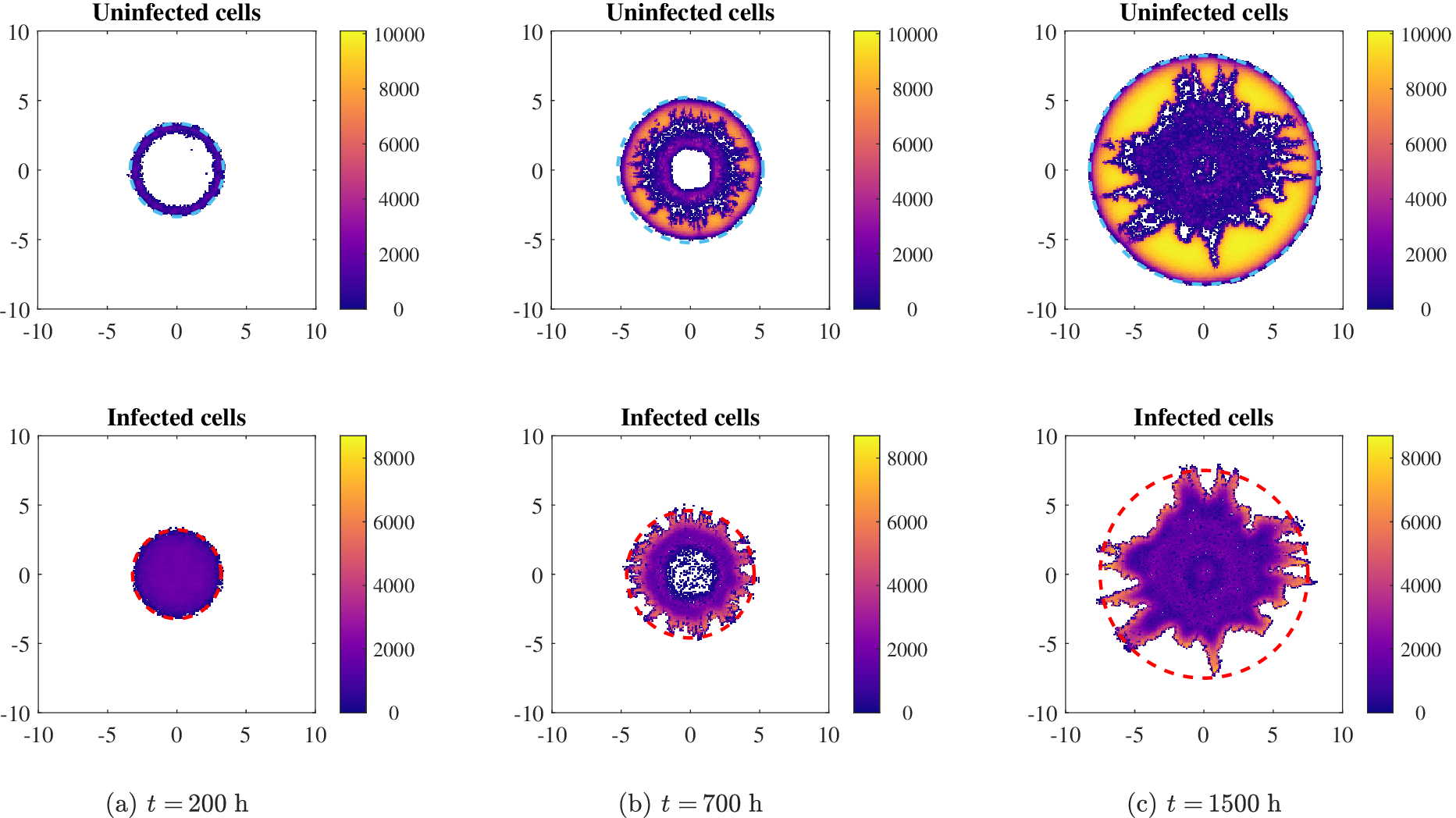}
		\caption{Numerical simulation of the discrete model with pressure-driven movement in two spatial dimensions at three different times. The parameters employed are the ones given in Table \ref{tab:parametersB1}, with the exception of the diffusion coefficient of viral particles $D_v$ (which is set to $10^{-5}\;$mm$^2$/h), the death rate of infected cells $q$ (which is set to $8.33\times 10^{-3}\;$h$^{-1}$, i.e. one-fifth of the reference values) and the initial radius of viral infection $R_v$ (which is set equal to $R_u$); furthermore, we use the values $\alpha=3500\;$viruses/cells and $q_v=1\;$h$^{-1}$. The dashed cyan circles in the top panels represent the expected positions of the uninfected invasion fronts in the absence of treatment, travelling at speed $\sqrt{\rev{D_P} p/2}$. The dashed red circles in the bottom panels represent the front of the infected cells given by the numerical solution of Eq. \eqref{eq:pv}.}
		\label{fig:pv_fingers}
	\end{figure}

	Reasonable increases of the effective infection rate $\btwo$ do not lead to a better infiltration of the infection, as this causes a decrease in central cell density and creates the need for infected cells to move against a pressure gradient. Also, as we will show shortly, oscillations emerge in some parameter ranges associated with an effective infection, but are suppressed at low values of viral diffusion. On the other hand, building upon the results of Ref. \cite{morselli23}, we expect that decreasing the death rate of infected cells $q$ leads to a better outcome.

	Fig. \ref{fig:pv_fingers}, along with the video accompanying it (see electronic supplementary material S6), indeed shows that a higher value of viral burst size $\alpha$ and lower values of viral decay $q_v$ and death rate of infected cells $q$ allow the infection to propagate for several hundreds of hours; nevertheless, stochastic fluctuations lead to local increases of uninfected cells, which in turn hinder the propagation of the infection in some directions. Observe how the stochastic events stopping the infection take place at different times in different locations, giving rise to interesting finger-shaped structures. It is also of notable interest to see that some areas of infection may be temporarily disconnected by the rest of the infection front (Fig. \ref{fig:pv_fingers}b-c). Overall, the infection is never completely halted, as viral diffusion always allows for some additional movement; this however does not lead to a treatment success.
	
	In general, the most common outcomes are either a complete confinement of the infection (as in Fig. \ref{fig:Dvlow}c) or a good agreement between the discrete and the continuum model (not shown here, as the parameter values are not biologically relevant). Stochastic patterns similar to the one of Fig. \ref{fig:pv_fingers} appear in some specific parameter ranges, which constitute an intermediate situation between the two ones just discussed.

	\subsection{Significant treatment success: emergence of oscillations}
	\label{sec:oscillations}
	
\begin{figure}
		\centering
		\includegraphics[width=\linewidth]{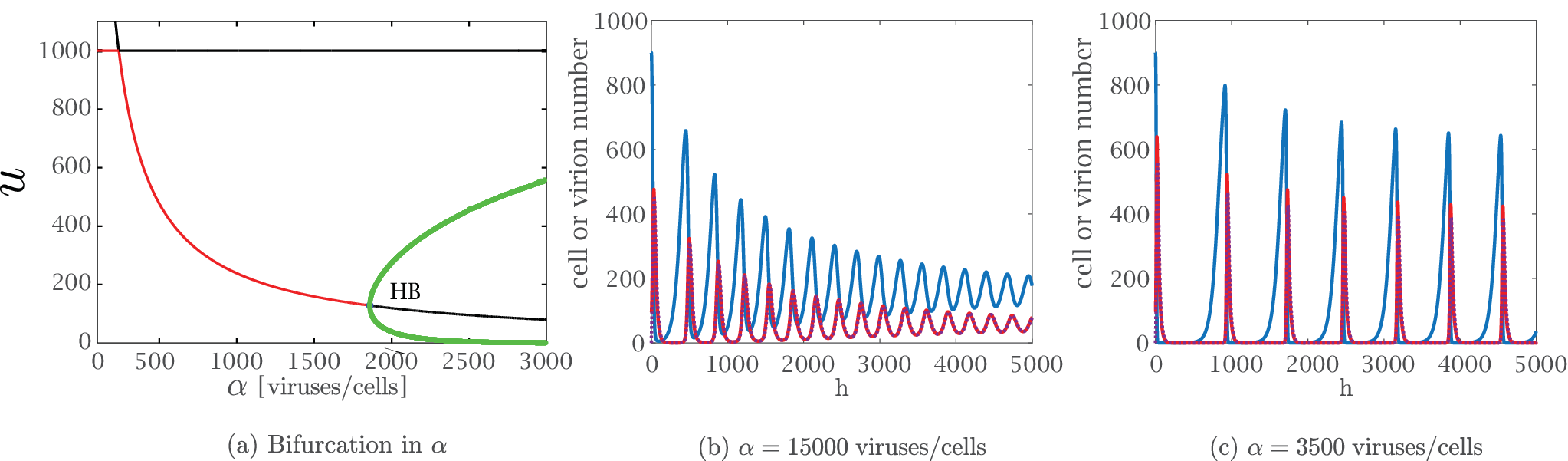}
		\caption[Bifurcation of the ODE for the infections with viral dynamics]{(a) One parameter bifurcation plot in $\alpha$ of Eq. \eqref{eq:odev}, with other parameters as in Table \ref{tab:parametersB1} and $q_v=0.167\,$h$^{-1}$. The green dots show the maximum and minimum values of $u$ for the stable limit cycle oscillations. The solid lines show the equilibrium value of $u$; the line is red if the equilibrium is stable and black if it is unstable. The Hopf bifurcation is denoted by HB.  Observe that for low values of $\alpha$, the infection-free equilibrium close to carrying capacity is stable.
		(b-c) Numerical simulation of Eq. \eqref{eq:ode} with the parameters as in Table \ref{tab:parametersB1} and different values of the viral burst size $\alpha$. The solid blue lines represent uninfected cells, the solid red lines infected cells and the dotted purple lines represent viral density multiplied by $q_v/(\alpha q)$; the two latter quantities superimpose almost exactly. The oscillations become wider as $\alpha$ increases, in accordance with the bifurcation diagram of panel (a).}
		\label{fig:bif}
	\end{figure}

\begin{figure}
		\centering
		\includegraphics[width=\linewidth]{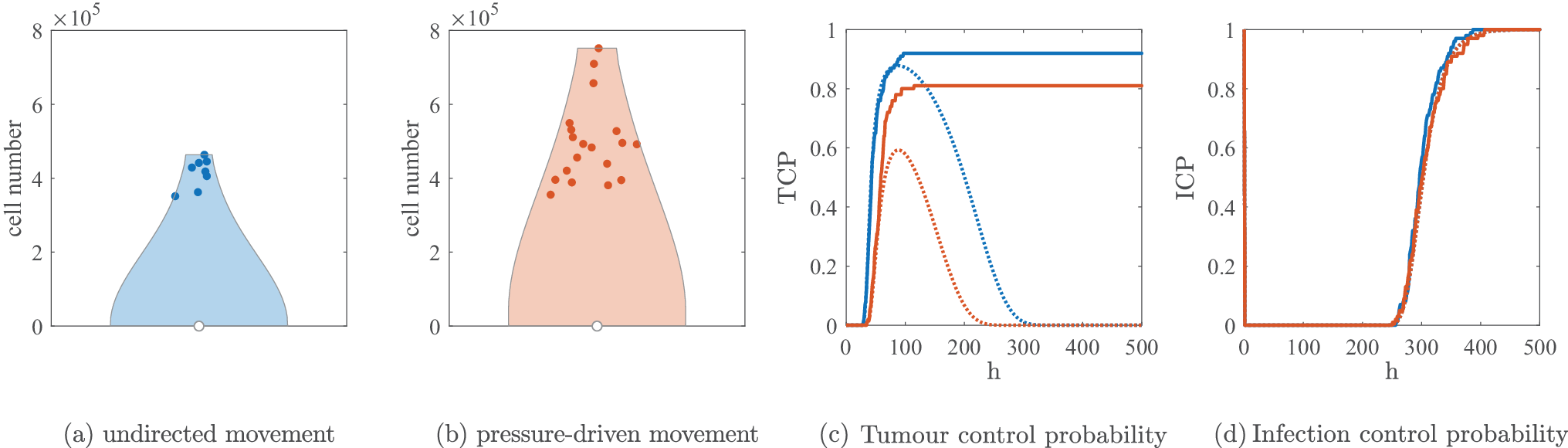}
		\caption{(a)-(b) Violin plots of the final cell number at time $t = 1500\;$h obtained from one hundred simulations	of the agent-based models with the parameters given in Table \ref{tab:parametersB1}, $\alpha=3500\;$viruses/cells and $q_v=8.33\times 10^{-2}\;$h$^{-1}$; the latter has been set to half of the lowest reference value to model a more effective viral infection. We also consider $R_v=R_u$ in the initial conditions to model a wide initial infection. The blue dots show the single results greater than zero. The white dots show the median, which is zero in both cases. (c)--(d) Tumour and infection control probabilities for the two agent-based models from Eqs. \eqref{eq:tcp_dis} and Eq. \eqref{eq:icp_dis} (undirected movement in solid blue and pressure-driven movement in solid red) and their corresponding continuous counterparts via the Poissonian assumption from Eqs. \eqref{eq:tcp_con} and Eq. \eqref{eq:icp_con} (same colours, with dotted lines).}
		\label{fig:violin}
	\end{figure}

	As we anticipated, it is well known that solutions of Eq. \eqref{eq:lv} may exhibit persistent oscillations for some parameter values; this is in line with the fact that the non-spatial model of Eq. \eqref{eq:eq_l} presents a Hopf bifurcation\rev{, as rigorously analysed in Refs. \cite{baabdulla23,bansod26}}. Fig. \ref{fig:bif}a shows a codimension one bifurcation plot in $\alpha$, whereas Fig. \ref{fig:bif}b-c shows the time series for three different values of $\alpha$.  We remark that a similar behaviour is also observed, for example, as $\bthree$ increases or as $q_v$ decreases.
	The one-dimensional setting is not ideal for studying oscillations, as local tumour extinctions due to stochasticity are very common: indeed, the typical behaviour observed in such parameter ranges is analogous to the one described below in relation to Fig. \ref{fig:violin}. We therefore concentrate our attention to two spatial dimensions: the higher number of cells involved allows us to describe oscillations, as well as stochastic-driven partial and complete eradication.
	
	\begin{figure}
		\centering
		\includegraphics[width=\linewidth]{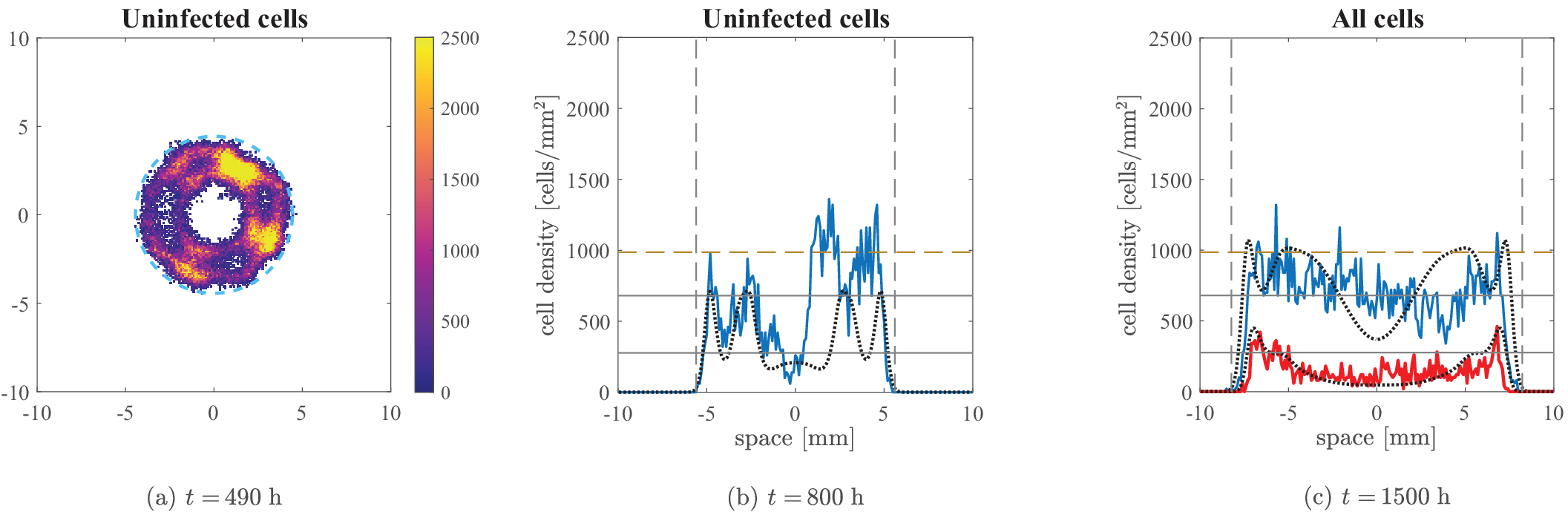}
		\caption[Two-dimensional simulations with viral dynamics, undirected cell movement, high viral burst size and low viral decay]{Numerical simulations of the discrete model with undirected movement in two spatial dimensions at three different times. The parameters employed are the ones given in Table \ref{tab:parametersB1}, $\alpha=3500\;$viruses/cells and $q_v=1.67\times 10^{-1}\;$h$^{-1}$. Panel (a) shows uninfected cell density from a single simulation to show the impact of stochasticity; the dashed cyan circle represents the expected positions of the tumour invasion front, travelling at speed $2\sqrt{\rev{D_U}p}$. Panels (b) and (c) represent cell densities on the horizontal section of the domain $[-L,L]\times\{0\}$, obtained by averaging five simulations: the density of the uninfected tumour cells is the blue solid line and the density of the infected tumour cells is the red solid line (not shown in panel (b) to avoid the superimposition with uninfected cell density); the viral density is not shown to improve readability. The solutions of Eq. \eqref{eq:lv} are represented with dotted black lines. The vertical black dashed lines represent the expected positions of the uninfected invasion front, travelling at speed $2\sqrt{\rev{D_U}p}$. The horizontal solid black lines show the equilibrium of the ODE given by Eq. \eqref{eq:eq_lv} and the horizontal dashed yellow line represents the expected uninfected density at the front given by Eq. \eqref{eq:ubar} (only relevant in panel (c)). The maximum of the axes and the colorbars were scaled to enhance readability and are much lower than the maximum over time of the quantity plotted.}
		\label{fig:lv_oscillations}
	\end{figure}

\begin{figure}
		\centering
		\includegraphics[width=\linewidth]{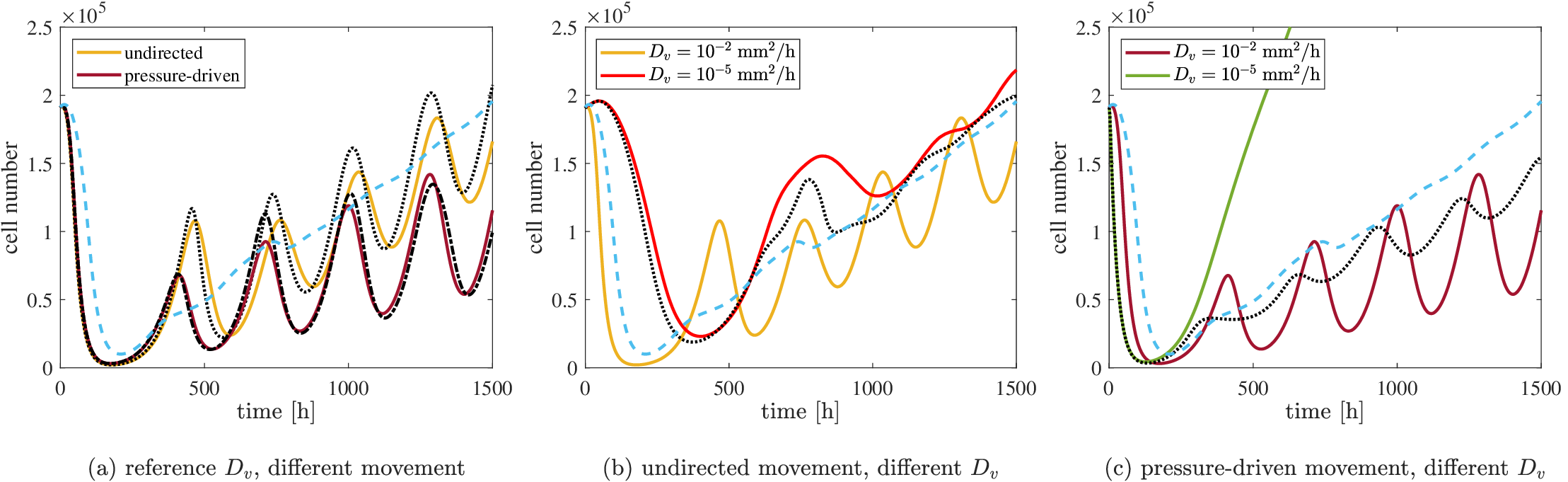}
		\caption[Comparison between oscillations in the sum of tumour cells in different scenarios]{Comparison between the sum of tumour cells in different scenarios. The parameters employed are the ones given in Table \ref{tab:parametersB1}, with $\alpha=3500\;$viruses/cells and $q_v=1.67\times 10^{-1}\;$h$^{-1}$; $D_v$ assumes different values in panels (b) and (c). Solid lines represent the average of five agent-based simulations in the case of undirected cell movement (yellow and red) and pressure-driven cell movement (purple and green). All the other lines represent the sum of cells obtained from continuous models. The dashed blue lines refer to the solution of Eq. \eqref{eq:l}; they are in a different colour than the other PDE solutions, to highlight the fact that it is related to a two-equations model, which does not exhibit persistent oscillations. The black lines refer to the solutions of either Eq. \eqref{eq:lv} or  Eq. \eqref{eq:pv}: in panels (b) and (c) we show for simplicity only the case of lower viral diffusion, as the reference cases are already shown in panel (a) and their qualitative behaviour is similar to the discrete models. }
		\label{fig:oscillations_sum}
	\end{figure}

	\paragraph{Undirected cell movement} Fig. \ref{fig:lv_oscillations}, along with the video accompanying it (see electronic supplementary material S7), shows an overall excellent quantitative agreement between numerical solutions of the system of PDEs \eqref{eq:lv} and the average over five numerical simulations of the corresponding agent-based model in two spatial dimensions. After the initial infection reaches the tumour boundary, the number of cancer cells decreases significantly; as a consequence, stochasticity becomes relevant and, in some regions, cells go extinct. Fig.~\ref{fig:lv_oscillations}a shows the inhomogeneous pattern of uninfected cells caused by this. Similar patterns would also arise in the two-dimensional non-radial continuum model under stochastic perturbations. As time elapses, the spatial inhomogeneities quickly disappear, and both models show similar persistent oscillations in the tumour centre around the equilibrium values (coherently with what the bifurcation diagram shows). A slight delay in the oscillations is the main difference between the two modelling approaches: this is easy to see from the total number of cells, depicted in Fig. \ref{fig:oscillations_sum}a. The invasion speed of the tumour in the surrounding area is still $2\sqrt{\rev{D_U}p}$ and the uninfected cell density at the front is approximately given by Eq. \eqref{eq:ubar}, despite some fluctuations. It is interesting to note that highly dynamic patterns may emerge when viral injections at multiple sites are considered \cite{baabdulla23}: this can be easily checked through VisualPDE \cite{walker23}, using the simulation available at the link \url{https://visualpde.com/sim/?mini=raZybpXG}.

	Fig. \ref{fig:oscillations_sum}a shows the comparison with the solution of Eq. \eqref{eq:l} to remark once again that oscillations cannot be obtained in the two-equations model even for high infection rates. Furthermore, Fig. \ref{fig:oscillations_sum}b illustrates that oscillations are almost non-existent for lower viral diffusion, resembling the behaviour of Eq. \eqref{eq:l}; this confirms once again the excellent quality of the quasi-steady assumption as spatial viral dynamics become negligible and the reactions involving the virus increase their rate.
	
	\begin{figure}
		\centering
		\includegraphics[width=\linewidth]{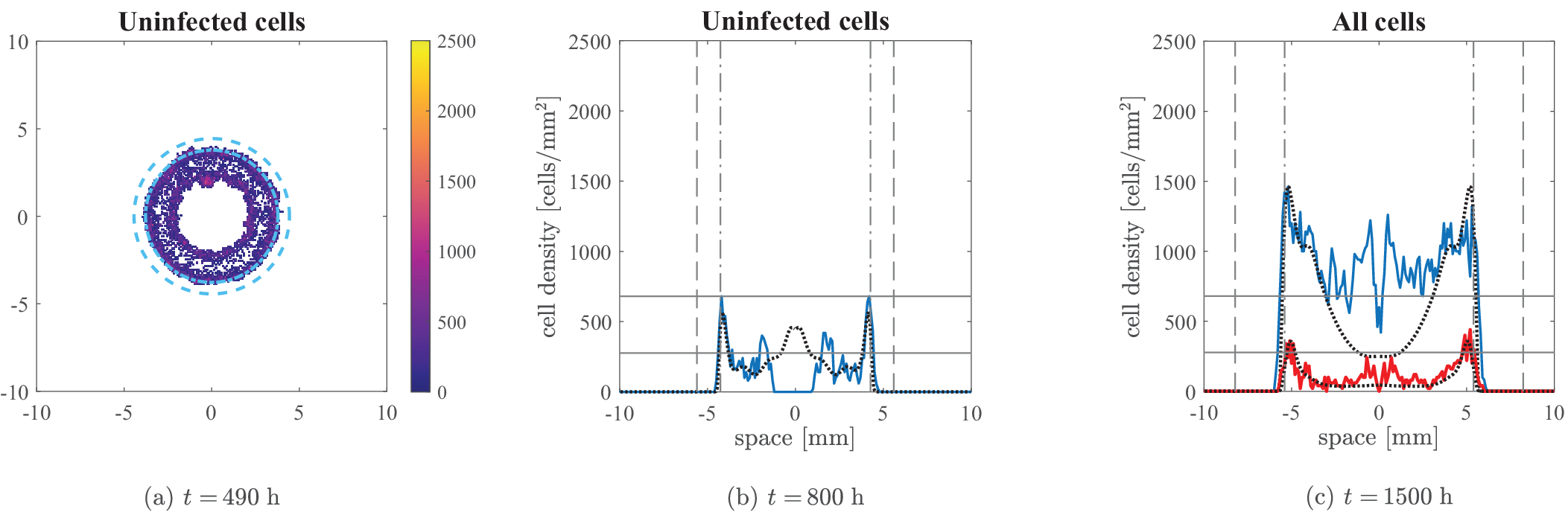}
		\caption[Two-dimensional simulations with viral dynamics, pressure-driven cell movement, high viral burst size and low viral decay]{Numerical simulations of the discrete model with pressure-driven movement in two spatial dimensions at three different times. The parameters employed are the ones given in Table \ref{tab:parametersB1}, $\alpha=3500\;$viruses/cells and $q_v=1.67\times 10^{-1}\;$h$^{-1}$. All the graphical elements have the same meaning as per Fig. \ref{fig:lv_oscillations}, with the vertical black dashed lines representing the expected positions of an untreated tumour, travelling at speed $\sqrt{\rev{D_P} p/2}$ and the vertical black dash-dotted lines representing the expected invasion speed in presence of virotherapy, given by Eq. \eqref{eq:speed_p_ubar}. Observe that in this second case we consider a wave starting from an initial radius of $3\;$mm rather than $R_u=2.6\;$mm, in order to compensate the transient behaviour in which the infection is far from the boundary of the tumour.}
		\label{fig:pv_oscillations}
	\end{figure}
	
	\paragraph{Pressure-driven movement} Fig. \ref{fig:pv_oscillations}a, along with the video accompanying it (see electronic supplementary material S8), shows that, as in the previous case, the number of cancer cells decreases significantly after the initial infection reaches the tumour boundary in both models and, in some regions, cells go extinct, originating inhomogeneous patterns. However, the subsequent regrowth does not allow a fast spatial spread due to the slow movement that cells experience when the pressure is low. Fig. \ref{fig:pv_oscillations}b indeed shows that uninfected cells take a long time in the agent-based model to move from the hollow ring into the central region; on the other hand, in the case of the PDE, the repopulation is much faster: there is never a complete extinction in the centre and the increase of cells' number in some areas is caused by regrowth rather than movement. This difference is particularly evident from the second peak in Fig. \ref{fig:oscillations_sum}a. The difference between the two models tends to disappear as time goes on, but Fig. \ref{fig:pv_oscillations}c still shows some dissimilarities.
	
	Both models have an invasion speed much lower than $\sqrt{\rev{D_P} p/2}$: indeed, the nonlinear diffusion is affected by the scale of the cell density, as explained in Appendix \ref{app:wave}. In our case, the oscillatory behaviour of the system does not allow to easily identify a unique, invariant speed. Fig. \ref{fig:pv_oscillations} shows that we can obtain a good approximation by setting $\bar{u}=1800\;$cells/mm$^2$, which corresponds to an average of the peak over time. The change in the speed in presence of effective infections is a significant difference from the case of undirected cell movement and, as a result, the overall cell numbers involved are lower in the second case (see Fig. \ref{fig:oscillations_sum}a). It is also interesting to remark that inhomogeneous infection patterns may result in an anisotropic spread of the tumour in the surrounding environment: we refer the interested reader to the VisualPDE simulation available at the link \url{https://visualpde.com/sim/?mini=e0LGK0st}.
	
	Let us remark once again that oscillations are only possible when the viral diffusion is high enough. If this is not the case, the behaviour of Eq. \eqref{eq:pv} is well approximated by Eq. \eqref{eq:p}, central infections remain constrained to the centre of the tumour and wide infections lead to a travelling wave without relevant oscillations\rev{. The} agent-based model results into a localised infection in all the relevant parameter ranges, as shown in Fig. \ref{fig:oscillations_sum}c.

	\paragraph{Stochastic extinction}
	The oscillations analysed so far are able to drive the tumour \rev{to} low densities, \rev{where} stochastic fluctuations may cause the death of all the agents in some regions; nevertheless, oscillations are not wide enough to lead the whole tumour close to extinction. It is important to see if it is possible to increase the effectiveness of the therapy and have the possibility of a full tumour remission. We therefore decrease the value of $q_v$ to half of the reference value, so that the oscillations can reach lower cell densities. We also consider an initial wide viral infection (i.e., $R_v=R_u$) to study the effect of a higher initial viral load. Fig. \ref{fig:violin}a-b summarise the cell numbers of one hundred simulations for both cell movement at time $t=1500\;$h. In both cases, the wide majority of the simulations shows the extinction of the tumour; the case of undirected cell movement appears more successful, with only eight simulations with a surviving tumour (in contrast to nineteen in the other case) and an overall lower cell density at the final time. It is convenient to track the distribution over time of the extinctions of the agent-based model through the \textit{tumour control probability}, defined as 
	\begin{equation}
		\label{eq:tcp_dis}
		\text{TCP}(\tau n)\coloneqq \frac{1}{M} \abs*{\Set{m |\sum_h U_h^n=0 \text{ in simulation } m}}
	\end{equation}
	where $M=100$ is the number of simulation performed and $\abs{\cdot}$ denotes the cardinality of the set. In other words, TCP$(\tau n)$ is the ratio of simulations that result in the extinction of the uninfected tumour cells in the interval $[0,\tau n]$. As infected cancer cells cannot proliferate, their survival does not affect the treatment outcome. Fig. \ref{fig:violin}c shows that all the extinctions happen within the first $200\;$h. A further insight is provided by the \textit{infection control probability}, defined as 
	\begin{equation}
		\label{eq:icp_dis}
		\text{ICP}(\tau n)\coloneqq \frac{1}{M} \abs*{\Set{m |\sum_h I_h^n=0 \text{ in simulation } m}}
	\end{equation}
	Fig. \ref{fig:violin}d shows the eradication of the infection in all the simulations performed. This explains the binary outcome that we observe: either all cancer cells become infected and eventually die out during the first oscillation, or a few cells avoid the infection and give rise to a recrudescence of the cancer.

	Continuum models cannot describe stochastic extinctions, but still provide meaningful information. In particular, the minimal cell number of the solution of Eq. \eqref{eq:lv} is lower than the analogous quantity for Eq. \eqref{eq:pv}, in line with the greater effectiveness of the therapy in the individual counterpart of Eq. \eqref{eq:lv}. A more quantitative comparison can be made through the \textit{Poissonian tumour control probability}
	\begin{equation}
		\label{eq:tcp_con}
		\text{TCP}(t)\coloneqq \exp\Bigl[- 2\pi\int_0^{R} u(t,r)\rev{r}\di r\Bigr]
	\end{equation}
	which is based on the assumption that the number of surviving uninfected cells follows a Poisson distribution. In Ref. \cite{gong13}, the Poissonian TCP defined as above was compared with more complex expressions and it was found to perform equally well in the case of radiotherapy. Fig. \ref{fig:violin}c shows an excellent quantitative agreement at early times between the expressions of Eqs. \eqref{eq:tcp_dis} and \eqref{eq:tcp_con}. The agreement is clearly lost as the continuum model shows the recrudescence of the tumour. It is also interesting to observe that the maximum value of the Poissonian TCP is slightly lower than the asymptotic value of the discrete TCP in both cases. A possible explanation for this discrepancy is that localised cell clusters are more effectively infected than wider, spatially homogeneous populations with a lower density; in the agent-based models, stochastic fluctuations during the infection naturally generate such small clusters, whereas the continuum model represents only averages and therefore cannot reproduce this effect. Nevertheless, the Poissonian TCP still provides a meaningful insight on the underlying stochastic model.
	
	We can follow the same approach to define the \textit{Poissonian infection control probability}
	\begin{equation}
		\label{eq:icp_con}
		\text{ICP}(t)\coloneqq \exp\Bigl[- 2\pi\int_0^{R} i(t,r)\rev{r}\di r\Bigr]
	\end{equation}
	Fig. \ref{fig:violin}d shows an excellent quantitative agreement between the expressions of Eqs. \eqref{eq:icp_dis} and \eqref{eq:icp_con}, meaning that the discrepancies mentioned above do not affect the dynamics of infected cells.

	Building upon the strong quantitative agreement observed so far, we now focus on the continuum model and examine how parameter variations influence the extinction probability. We adopt the same parameter settings as in the simulations shown in Fig.~\ref{fig:violin} and investigate how the viral burst size $\alpha$ affects the maximal value attained for positive times by $\text{TCP}(t)$ and $\text{ICP}(t)$ in the numerical solutions of Eqs.~\eqref{eq:lv} and \eqref{eq:pv}. Note that the maximal value of $\text{TCP}(t)$ corresponds to the probability of tumour extinction within the time window considered, since in the agent-based model no new cancer cells can arise after extinction has occurred. The situation is different in the case of $\text{ICP}(t)$, as an infection can emerge even in absence of infected cells. In particular, the initial conditions in Eq.~\eqref{eq:initialv} yield $\text{ICP}(0)=1$, which does not reflect the subsequent dynamics: the initial viral load inevitably triggers infection (see also Fig.~\ref{fig:violin}d). By contrast, extinction of infected cells at later times generally indicates complete eradication of the infection, due to the rapid decay of viral particles. For this reason, we restrict our analysis to $t>0$ when computing the maximal values of both probabilities over time.

\begin{figure}
		\centering
		\includegraphics[width=\linewidth]{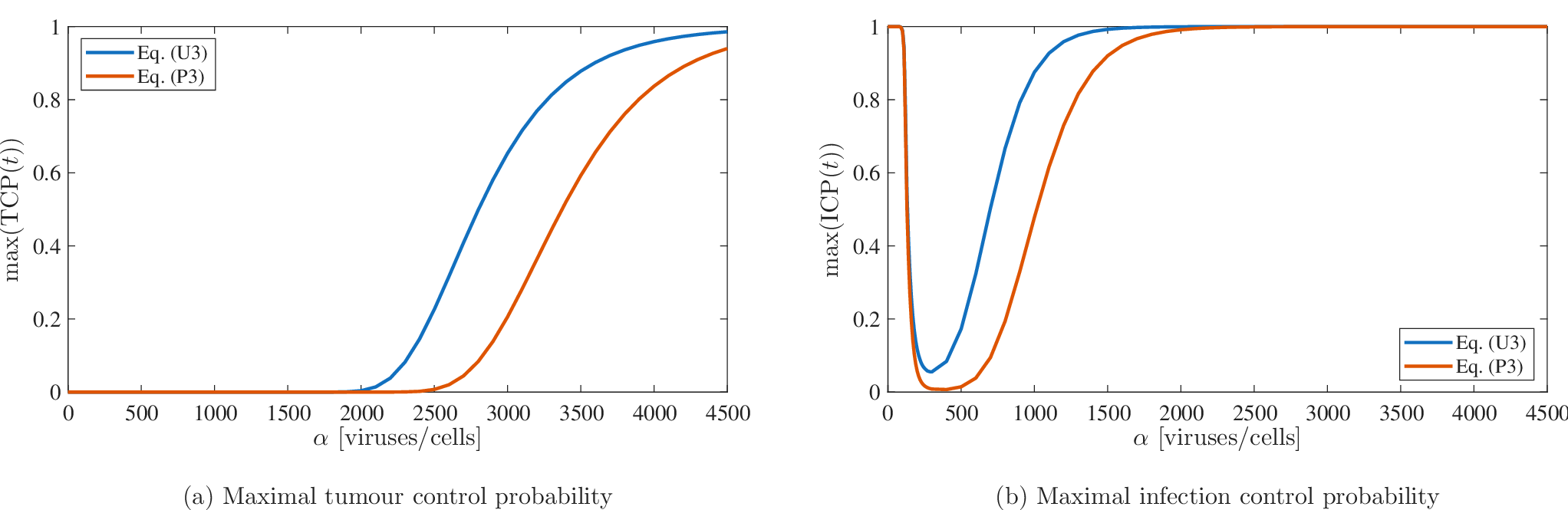}
		\caption{\rev{Maximal values attained at positive time by $\text{TCP}(t)$ and $\text{ICP}(t)$ for the numerical solutions of Eqs. \eqref{eq:lv} and \eqref{eq:pv} for different values of $\alpha$. The other parameters are the ones listed in Table \ref{tab:parametersB1}, with the exception of $q_v$ that is set to  $8.33\times 10^{-2}\;$h$^{-1}$ (i.e., half of the lowest reference value) to model a more effective viral infection. We also consider $R_v=R_u$ in the initial conditions to model a wide initial infection.}}
		\label{fig:tcp_extra}
	\end{figure}

	Fig. \ref{fig:tcp_extra} shows that the tumour extinction has a higher probability as $\alpha$ increases, confirming once again that a higher viral burst is associated to a more effective treatment. For $\alpha$ high enough the probability is very close to one, meaning that in practice we can always expect tumour extinction; such values however are probably beyond the biologically meaningful range. It is also interesting to observe that the eradication of the infection is much more common and could be associated either to the eradication of the full tumour (for high values of $\alpha$) or to the total failure of the treatment (for low values of $\alpha$).
	
	Our results show that, at least in principle, virotherapy may cause a complete tumour remission if the viral infection is extremely effective; this is interesting, but it remains an open question how to translate \rev{it} in a clinical settings. We remark that, for such low cell densities, the immune clearance of the surviving cells could facilitate tumour extinction; on the other hand, an early immune response may hinder the infection and lead to an overall decrease of the therapy effectiveness, as discussed in Ref. \cite{morselli24}.

\section{Conclusions}
\label{sec:conclusion}
\begin{table}
		\centering{
			\footnotesize
			\begin{tabular}{cp{0.35\linewidth}p{0.35\linewidth}}
				\toprule
				& \textbf{Undirected cell movement} & \textbf{Pressure-driven cell  movement} \\
				\midrule
				\textbf{Quasi-steady virus} 
				& Continuum model: Eq. \eqref{eq:l} & Continuum model: Eq. \eqref{eq:p} \\
				& Comprehensive theoretical understanding & Partial theoretical understanding\\
				& Constant tumour invasion speed & Slower tumour invasion at low cell densities \\
				& Central infections infect all the tumour & Central infections cannot spread \\
				& No significant stochastic effects & Significant stochastic effects \\
				& No oscillatory behaviour & No oscillatory behaviour
				\\
				\midrule
				\textbf{Explicit virus} 
				& Continuum model: Eq. \eqref{eq:lv} & Continuum model: Eq. \eqref{eq:pv} \\
				& Comprehensive theoretical understanding & Partial theoretical understanding \\
				& Constant tumour invasion speed & Slower tumour invasion at low cell densities \\
				& Central infections infect all the tumour (with a speed that may be different than in case of Eq. \eqref{eq:l}) & Central infections only spread with high $D_v$ (with a speed that may be different than in case of Eq. \eqref{eq:l})\\
				& Significant stochastic effects in presence of oscillations that lead to partial or complete extinction & Significant stochastic effects for low $D_v$ and in presence of oscillations that lead to partial or complete extinction\\
				\bottomrule
			\end{tabular}
		}
		\caption{Comparison of tumour invasion models with implicit vs explicit virus formulation and different cell movement assumptions.}
		\label{tab:comparison}
	\end{table}

	In this work, we have investigated the behaviour of several models of oncolytic virotherapy when viral dynamics is explicitly formulated in the equations. A comparison with known and new treatment scenarios for systems that do not involve such mathematical modelling has highlighted some valuable ideas and also improved our knowledge of the role of motility of single and collective tumour cell(s) when a virus is present. The results are summarised in Table \ref{tab:comparison}.
	
	For explicit models, the rules governing the spatial movement of viruses appear to affect the infection more significantly than the rules governing tumour cell movement. Nonetheless, cell movement does still retain its influence on tumour invasion, especially when movement is constrained: only pressure-driven equations describe a slower invasion for low tumour cell numbers. This is a relevant aspect in the overall success or failure of the therapy and bears clinical consequences that are important when analysing patient's results. Overall, undirected movement (with or without viral diffusion) and pressure-driven movement with viral diffusion exhibit similar behaviour in regards to the spread of the infection throughout a tumour. The main difference among all models is the speed of the infections and, when this is crucial for applications, our result suggest that the viral population should be openly included in the formulation. An additional difference that has therapeutic value is the appearance of oscillations between viral and tumour populations, which is not a feature shared by all models. For instance, a pressure-driven tumour not explicitly infected by a viral population (or with a low viral diffusion coefficient) exhibits a very different behaviour than a model with unhindered tumour diffusion. Notably, the presence of a viral population can also lead to persistent oscillations, which, as we have seen, can significantly affect the overall outcome in some cases. This could be very valuable when considering, for example, the effect of viral vectors on outcomes for some chronic blood tumours, which have been known for many years to exhibit semi-regular oscillatory regimes \cite{CHIKKAPPA1980Blood, RODRIGUEZ1976AMJ, Vodopick1972NEJM}.
	
	Although undirected movement without a viral population can be a suitable modelling assumption in some cases, our findings show that the choice of pressure-driven movement is often more appropriate for realistic outcomes and it can reproduce a wider range of phenomena. Importantly, stochasticity mostly plays a notable role for a pressure-driven tumour in the absence of an explicit viral infection or with a sufficiently low viral diffusion coefficient. As shown, some additional scenarios can also emerge when strong oscillations are present and chance can drive the systems to a  partial or complete extinction outcome. In these regards, a two-equation model such as Eq. \eqref{eq:l} is a good predictor for outcomes, with discrepancies arising mainly because of the presence of oscillations (for both types of tumour movements) or for factors that tend to spatially block the infection when pressure-driven cases are considered.
	
	Furthermore, let us emphasise once more that undirected cell movement, which corresponds to standard diffusion, is the most elementary approach and does not consider any external influence. This has an important consequence for the characterisation of infection waves, for which that model gives only a basic interpretation and often incomplete. The simulations discussed in this work confirm that the invasion speed of a tumour mass in the surrounding environment is seen as constant (irrespectively of the height of the infection front) and a viral load localised in the centre of the cancer faces no obstacle in spreading in the whole domain through cells' displacement. On the other hand, pressure-driven movement, which corresponds to nonlinear diffusion, relies on the assumption that a cell only moves because of the pressure exerted by the surrounding cells: as a result, movement is directed towards less crowded areas and is more relevant for steep changes in the cell density. This last approach is well-justified by biological evidence and can model the decreased invasion speed that results from lower cell densities. 
	
	In conclusion, it is important to remark that, for pressure-driven models, cells' displacement alone cannot spread viral infection and central infections remain localised. As previously observed, even wide infections may be stopped by stochastic events~\cite{morselli23}, but the situation significantly changes when viral particles are allowed to freely move around the tumour and cell-to-cell contact becomes less relevant for the infection. In that case, our theoretical and computational results have explained that this infection is qualitatively similar to the case of undirected cell movement. We can conclude that pressure-driven movement allows us to describe both the situations of constrained and unconstrained cellular and viral movements, while such constraints cannot be modelled with undirected movement. The increased realism of pressure-driven movement, though, comes at the cost of a lower degree of theoretical understanding of the models.
	
	Finally, a couple of possible extensions of this work are worth being mentioned. Firstly, it would be interesting to include actual immunotherapies and other contemporary clinical therapies in the explicit models presented here. This could allow to extend the theoretical and computational predictions we have discussed and compare them with more realistic scenarios. The evaluation of the role of spatial dependency for viral infection could also \rev{be} particularly interesting. For example, looking at how tumour shape influences the outcome of virotherapy or where and how the way viral load is injected can give rise to different infection speeds are both valuable ideas for future works.

\appendix

\section{Speed of the travelling waves}
		\label{app:wave}
		
		\subsection{Uninfected cells}
		\label{app:wave_u}
		
		\paragraph{Undirected movement}
		Let us first consider the equation
		\[
		\pt u=\rev{D_U}\pxx u +p\Bigl(1-\frac{u}{K}\Bigr) u
		\]
		We can define the nondimensional variables 
		\[
		C=\frac{u}{K},\quad \hat{t}=p t, \quad \hat{x}= x \sqrt{\frac{p}{\rev{D_U}}}
		\]
		and obtain the equation
		\begin{equation}
			\label{eq:adim_l}
			\partial_{\hat{t}} C=\partial_{\hat{x}\hat{x}}^2 C+(1-C)C
		\end{equation}
		
		As it is well-known, this equation admits as solutions positive travelling waves of the form $C(\hat{x},\hat{t})=\varphi(\hat{x}-c\hat{t})$ connecting the two equilibria if $c\geq c^*\coloneqq2$; an initial condition with compact support evolves into a wave that travels with minimal speed \cite{fisher37,kolmogorov37}. The dimensional value of the minimum speed is $2\sqrt{\rev{D_U}p}$.
		
		Let us remark that the speed is independent of the value of the carrying capacity: in other words, if the reaction part of the equation is such that the equilibrium is at $u=\bar{u}$ instead of $u=K$ (as in the case of Eq. \eqref{eq:lv} when the infection is present in the whole tumour), then the adimensionalisation $C=U/\bar{u}$ keeps the speed unchanged. This is in line with our result of Fig. \ref{fig:lv_oscillations}, which shows that the speed value is constant even in the case of an extremely effective infection.
		
		\paragraph{Pressure-driven movement}
		Let us now consider the equation
		\[
		\pt u=\frac{\rev{D_P}}{K}\px(u\px u) +p\Bigl(1-\frac{u}{K}\Bigr) u
		\]
		We can define the same nondimensional variables as before to obtain the equation
		\begin{equation}
			\label{eq:adim_p}
			\partial_{\hat{t}} C=\partial_{\hat{x}}(C\partial_{\hat{x}}  C)+(1-C)C
		\end{equation}
		Results analogous to the previous case still hold, but this time the minimum speed is $c^*\coloneqq 1/\sqrt{2}$ \cite{aronson80,newman80}. Its dimensional value is $\sqrt{\rev{D_P} p/2}$. It is straightforward to check that we recover the same speed value as in Eq. \eqref{eq:adim_l} by multiplying the diffusion coefficient by a factor of $8$: this is indeed the procedure that we follow whenever we compare models with different rules of movement, so that we avoid any bias due to the different propagation speed.
		
		As in the previous case, we are now interested in the situation of an equilibrium at the value $u=\bar{u}$ instead of $u=K$ (as in the case of Eq. \eqref{eq:pv} when the infection is present in the whole tumour). This requires a slightly different adimensionalisation: indeed, the diffusion coefficient is still $D/K$ and so we need to use the variables
		\[
		C=\frac{u}{\bar{u}},\quad \hat{t}=p t, \quad \hat{x}= x \sqrt{\frac{pK}{\rev{D_P}\bar{u}}}
		\]
		As a consequence, the dimensional value of $c^*$ is now
		\begin{equation}
			\label{eq:speed_p_ubar}
			\sqrt{\frac{\rev{D_P} p}{2}\ \frac{\bar{u}}{K}}
		\end{equation}
		meaning that a lower peak at the front is associated with a slower wave. 
		
		It is important to remark that we are not performing a rigorous analysis, which should take into account the exact form of the interactions between all the populations. Nevertheless, the speed obtained in Eq. \ref{eq:speed_p_ubar} is confirmed by our numerical results (see Fig. \ref{fig:pv_oscillations}).

		\subsection{Infected cells and virus}
		\label{app:wave_i}
		
		We now restrict our attention to undirected movement, as linearisation techniques do not provide insightful results about pressure-driven movement.
		
		\paragraph{Invasion into an uninfected region at carrying capacity (following Ref. \cite{baabdulla23})}
		
		\begin{figure}
			\centering
			\includegraphics[width=\linewidth]{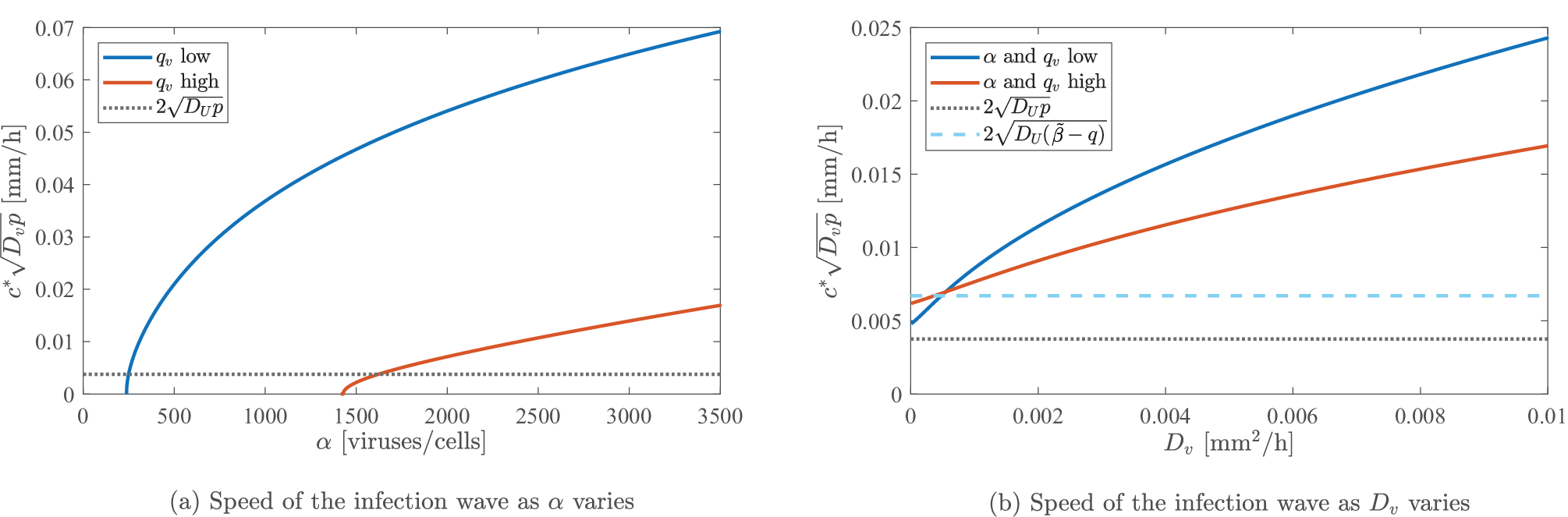}
			\caption{\rev{Theoretical speed of the travelling wave of viral infection invading an uninfected tumour at carrying capacity for different parameter values compared to the speed of uninfected front (black dotted line) and the speed of infection for Eq. \eqref{eq:l} (cyan dashed line). In panel (a), the value of $\btwo$ is not constant, so the comparison with Eq. \eqref{eq:l} is not particularly meaningful. Observe that for low values of $\alpha$ there is no travelling wave, as $i^*$ is negative and the equilibrium $(u^*, i^*)$ is unstable (see also Fig. \ref{fig:bif}).}}
			\label{fig:speed_extra}
		\end{figure}
		
		After adimensionalisation, Eq. \eqref{eq:lv} becomes
		\begin{equation}
			\label{eq:lv_adimensional}
			\begin{cases} 
				\partial_{\hat{t}} C&= D_{c} \partial_{\hat{x}\hat{x}}^2 C (1-C-I) - CV \\ 
				\partial_{\hat{t}} I &= D_{c} \partial_{\hat{x}\hat{x}}^2 I +CV- a I \\ 
				\partial_{\hat{t}} V&= \partial_{\hat{x}\hat{x}}^2 V+ \theta I- \gamma V 
			\end{cases} 
		\end{equation}
		with the nondimensional variables
		\[
		C=\frac{u}{K},\quad I=\frac{i}{K},\quad {V}= \frac{\beta }{K p} v,\quad \hat{t}=p t, \quad \hat{x}= x \sqrt{\frac{p}{D_{v}}}
		\]
		and parameters
		\[
		D_{c}=\frac{\rev{D_U}}{D_{v}},  \quad a=\frac{q }{p },\quad \theta =\frac{\alpha\beta q}{p^{2}}, \quad \gamma = \frac{q_v}{p}
		\]
		
		We then make an explicit ansatz of an exponentially decaying self-similar wave solution, namely
		\[
		(C(\hat{x},\hat{t}),I(\hat{x},\hat{t}),V(\hat{x},\hat{t}))= \left( \epsilon _{c} \, e^{-\lambda z}, \epsilon _{i} \, e^{-\lambda z}, \epsilon _{v} \, e^{-\lambda z} \right) , \quad \text{ where } \quad z=\hat{x}-c\hat{t}
		\]
		Substituting this in Eq. \eqref{eq:lv_adimensional} and considering only the linear part around $(1,0,0)$, we obtain
		\[
		\begin{bmatrix} D_{c} \lambda ^2 - c\lambda -1 &{} -1 &{} -1 \\ 0 &{} D_{c} \lambda ^2 - c\lambda - a &{} 1 \\ 0 &{} \theta &{} \lambda ^2 - c\lambda - \gamma \end{bmatrix} \begin{bmatrix} \epsilon _{c} \\ \epsilon _{i} \\ \epsilon _{v} \end{bmatrix} =\begin{bmatrix} 0 \\ 0 \\ 0 \end{bmatrix}
		\]
		From the characteristic equation of the matrix, we obtain that the previous system admits a non-trivial solution only if
		\[
		\Phi (\lambda,c) \coloneqq D_{c} \lambda ^{4} - c (\rev{D_c}+1) \lambda ^{3} + (c^{2}-D_{c}\gamma -a) \lambda ^{2} + c (\gamma + a) \lambda + a \gamma -\theta=0
		\]

		Using implicit differentiation, we can compute
		\[
		c'(\lambda ) =\frac{\Psi (\lambda ;c)}{2 c \lambda ^2-(\rev{D_c}+1) \lambda ^{3}+(\gamma +a) \lambda }
		\]
		where
		\[
		\Psi (\lambda ,c)\coloneqq 2 \lambda c^{2}+(\gamma +a-3(D_{c}+1)\lambda ^{2}) c+ 4\rev{D_c} \lambda ^{3}-2(D_{c} \gamma +a) \lambda
		\]
		Hence the minimum wave speed $c^*$ arises at the intersection of the two manifolds $\{\Phi (\lambda ,c)=0\} \cap \{\Psi (\lambda ,c)=0\}$. The value of $c^*$ can be found numerically, using the built-in \textsc{Matlab} function \texttt{fsolve}. The dimensional value of the speed is then $c^*\sqrt{D_v p}$\rev{, which is shown in Fig. \ref{fig:speed_extra} for different parameter values. Observe that higher values of $\alpha$ and lower values of $q_v$ result in a higher value of $\btwo$, corresponding to a more effective infection (in line with the results of Section \ref{sec:waves_sim}). Similarly, the infection propagates faster for high values of $D_v$, although equilibrium values are not affected.}

        \begin{figure}
			\centering
			\includegraphics[width=0.5\linewidth]{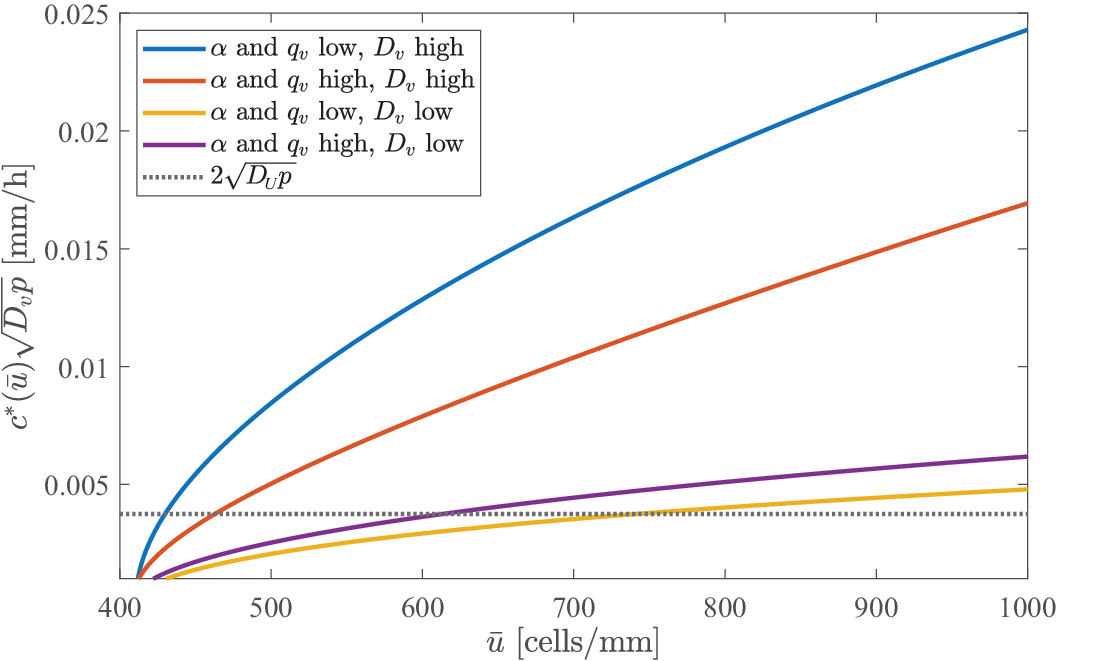}
			\caption{Dimensional minimum speed for different values of $\bar{u}$ compared to the speed of the uninfected travelling wave $2\sqrt{\rev{D_U}p}$. The parameters employed are the ones given in Table \ref{tab:parametersB1}, with both values of $\alpha$ and $q_v$. The higher value of $D_v$ is the reference one, while the lower value is $D_v=10^{-5}\;$mm$^2$/h.}
			\label{fig:ubar}
		\end{figure}

		\paragraph{Invasion into an uninfected region of constant density $\bar{u}$}

		The linearisation of the last two equations of Eq. \eqref{eq:lv_adimensional} around $(0,0)$ assuming $C=\bar{u}$ leads to 
		\[
		\begin{bmatrix} &{} D_{c} \lambda ^2 - c\lambda - a &{} \bar{u} \\ &{} \theta &{} \lambda ^2 - c\lambda - \gamma \end{bmatrix} \begin{bmatrix} \epsilon _{i} \\ \epsilon _{v} \end{bmatrix} =\begin{bmatrix} 0 \\ 0 \end{bmatrix}
		\]
		which admits a non-trivial solution only if
		\[
		\tilde{\Phi} (\lambda,c) \coloneqq D_{c} \lambda ^{4} - c (\rev{D_c}+1) \lambda ^{3} + (c^{2}-D_{c}\gamma -a) \lambda ^{2} + c (\gamma + a) \lambda + a \gamma -\theta \bar{u}=0
		\]
		The only difference with respect to the previous case is the term $-\theta \bar{u}$ that replaces $-\theta$. The expression of $c'(\lambda)$ remains unchanged and so does $\Psi(c,\lambda)$. The minimum wave speed $c^*(\bar{u})$ arises at the intersection of the two manifolds $\{\tilde{\Phi} (\lambda ,c)=0\} \cap \{\Psi (\lambda ,c)=0\}$.
		
		The expected peak of uninfected cells at the invasion front can be found by numerically solving for $\bar{u}$ the equation
		\[
		c^*(\bar{u})\sqrt{D_v p}=2\sqrt{D_U p}
		\]
		Fig. \ref{fig:ubar} shows the two curves for different values of the parameters; the values of $\bar{u}$ at the intersection are the ones used in Fig. \ref{fig:lv_low} (parameters used for the blue line), supplementary material S3 (parameters used for the red line), Fig. \ref{fig:Dvlow}a (parameters used for the yellow line) and Fig. \ref{fig:Dvlow}b (parameters used for the purple line). 

\section*{Supplementary information}
	
The supplementary material is available in the online version of the published paper, as well as at the link \url{https://doi.org/10.5281/zenodo.18877438}.

\bibliographystyle{siam}
\bibliography{bibliography}
	
\end{document}